AN ABSTRACT OF A THESIS

NOVEL HIGH EFFICIENCY QUADRUPLE JUNCTION SOLAR CELL WITH CURRENT MATCHING AND OPTIMIZED QUANTUM EFFICIENCY

Mohammad Jobayer Hossain

Master of Science in Electrical and Computer Engineering

A high photon to electricity conversion efficiency of 47.2082% was achieved by a novel combination of $In_{0.51}Ga_{0.49}P$, GaAs, $In_{0.24}Ga_{0.76}As$ and $In_{0.19}Ga_{0.81}Sb$ subcell layers in a quadruple junction solar cell design. The electronic bandgap of these materials are 1.9 eV, 1.42 eV, 1.08 eV and 0.55 eV respectively. This novel III-V arrangement enables the cell to absorb photons from the ultraviolet to deep infrared wavelengths of the solar spectrum. After careful consideration of important semiconductor parameters such as thicknesses of emitter and base layers, doping concentrations, diffusion lengths, minority carrier lifetimes and surface recombination velocities an optimized quadruple junction design has been suggested. Current matching of the subcell layers was ensured to obtain maximum efficiency from the proposed design. The short-circuit current density, open circuit voltage and fill factor of the solar cell are 14.7 mA/cm$^2$, 3.3731 V and 0.9553 respectively. In the design process, 1 sun AM1.5 global solar spectrum was considered.

The cell performance was also investigated for extraterrestrial illumination (AM0). A modified design is proposed for space applications. With a short circuit current density of 18.5 mA/cm$^2$, open circuit voltage of 3.4104 and the fill factor of 0.9557, the power conversion efficiency of the modified quadruple junction design is 44.5473% in space.

**Novel High Efficiency Quadruple Junction Solar Cell with Current Matching and**

**Optimized Quantum Efficiency**

___________________________________

A Thesis

Presented to

The Faculty of the College of Graduate Studies

Tennessee Technological University

by

Mohammad Jobayer Hossain

___________________________________

In Partial Fulfillment

of the Requirements for the Degree

MASTER OF SCIENCE

Electrical and Computer Engineering

___________________________________

December 2016

**CERTIFICATE OF APPROVAL OF THESIS**

**NOVEL HIGH EFFICIENCY QUADRUPLE JUNCTION SOLAR CELL WITH**

**CURRENT MATCHING AND OPTIMIZED QUANTUM EFFICIENCY**

by

Mohammad Jobayer Hossain

Graduate Advisory Committee:

\_\_\_\_\_\_\_\_\_\_\_\_\_\_\_\_\_\_\_\_\_\_\_\_\_\_       \_\_\_\_\_\_\_\_\_\_
Indranil Bhattacharya, Chairperson         Date

\_\_\_\_\_\_\_\_\_\_\_\_\_\_\_\_\_\_\_\_\_\_\_\_\_\_       \_\_\_\_\_\_\_\_\_\_
Satish Mahajan                                          Date

\_\_\_\_\_\_\_\_\_\_\_\_\_\_\_\_\_\_\_\_\_\_\_\_\_\_       \_\_\_\_\_\_\_\_\_\_
Ghadir Radman                                         Date

Approved for the Faculty:

\_\_\_\_\_\_\_\_\_\_\_\_\_\_\_\_\_\_\_\_\_\_\_\_\_\_\_\_
Mark Stephens
Dean
College of Graduate Studies

\_\_\_\_\_\_\_\_\_\_\_\_\_\_\_\_\_\_\_\_\_\_\_\_\_\_\_\_
Date

**DEDICATION**

This thesis has been dedicated to the amazing teachers of my life:

Zobaida Begum, my mother

Md. Jakir Hossain, my father

Abdul Karim Sarker, my grandfather

Md. Nur Islam, headmaster, Rajarhat Model Primary School, Kurigram, Bangladesh

Nazir Hossain, former B.Sc. teacher, Rajarhat Pilot High School, Kurigram, Bangladesh

Prof. Ratan Kumar Dev, former head, Math Department, Rangpur Govt. College, Bangladesh

# ACKNOWLEDGEMENT

I would like to specially acknowledge and thank Dr. Indranil Bhattacharya, the chairperson of my MS thesis committee, for giving me an opportunity to work on designing high efficiency solar cells— an area I was very much interested in and I decided before joining Tennessee Tech University as a student. His supervision, guidance and moral support throughout the research, especially at the time of failure are that made this thesis successful. I would also like to express my gratitude to Dr. Satish Mahajan, and Dr. Ghadir Radman for being the members in my MS committee and their wise suggestions.

I am pleased to have a cooperative and cordial environment in Dr. Bhattacharya's SOLBAT research group. The co-research mates Bibek Tiwari, Jagadish Babu, Rani Penumaka and Behnaz Papari were always supportive. Their suggestions, constructive criticisms and warm manners kept me live all through the research work.

I would like to thank the Office of Research and Department of Electrical and Computer Engineering for providing financial assistance without which this research would not have been possible. I would also like to thank the TTU Center of Energy Systems Research (CESR) for financial assistance during summer semesters for continuing my research.

All through my staying in TTU and conducting research, my parents, relatives and villagers in Bangladesh were the true sources of my inspiration, who urged me 'make us proud' when I came to US for the first time. I cannot express my gratefulness to them in words.

# TABLE OF CONTENTS

# LIST OF TABLES

# LIST OF FIGURES

# CHAPTER 1

# INTRODUCTION TO SOLAR CELL RESEARCH

## 1.1    Introduction

Electricity has become an inseparable part of our modern society. All the sectors of human civilization, be it health care, agricultural management, household applications, business activities, industrial production facilities or research— everything is dependent on electrical energy. The demand of electricity is increasing rapidly with increase in population as well as economic activities in the society. This needs tremendous increase in energy generation to keep up with the pace of energy requirements. The traditional sources of electrical energy such as coal, oil and natural gas have limited reserve i.e. only 892 billion tons of coal, 186 trillion cubic meters of natural gas, and 1,688 billion barrels of crude oil all over the world according to BP Statistical Review of World Energy [1]. At the present rate of usage it is predicted that coal will be exhausted in 103 years, the oil reserve in 53 years and the natural gas in 54 years respectively [1, 2]. Many environmental problems in today's world, for example global warming, climate change, air pollution, acid rain etc. happens because of burning fossil fuels. The large extent of pollution occurring today may one day threaten the existence of life on earth. Renewable energy sources such as solar, biomass, geothermal, hydroelectric and wind power generation has emerged as potential alternatives to meet the energy demands and to address the above mentioned environmental concerns. As compared to fossil fuels, renewable energy sources are unlimited and available in nature [3]. Solar Energy has become the most rapidly growing renewable source because of its advantages over other renewable counterparts. Solar electricity generation has several advantages such as: can be produced anywhere, simple to install, silent operation, longer lifespan, lesser maintenance and no radioactive decay.

Solar energy is the most abundant renewable energy resource on earth. About 885 million terawatt-hours (TWh) of solar energy reaches the earth's surface every year, which is 6,200 times greater than the energy consumed by mankind in 2008 and 4200 times the energy human civilization would need in 2035 [4]. The energy we receive from sun in just 1 hour and 25 minutes is sufficient to meet the energy requirements of the world for one year. Despite this huge potential, only a small portion of solar energy is utilized to generate electricity. Only 10% of the total energy supplied in the United States comes from renewable energy sources, and only half of that 10% comes from solar power [5]. Although there is no fuel cost, the installation cost is still high. In contrast fossil fuels are still cheaper. Currently, electricity production from renewable sources costs $24.34/MWh where it costs only $0.44 /MWh for electricity from fossil fuels [5]. Government incentives and public awareness has led to the installation of renewable energy generation infrastructures that is predicted to generate 40% of the total electricity in the USA by 2030, according to the Department of Energy. Photovoltaic industry is projected to be a $345.59 billion industry by 2020 [6].

Most of the commercially available solar cells are single junction silicon solar cells. It has been dominating the solar industry since its inception because of the abundance in silica. Silicon (Si) has an indirect bandgap of 1.14 eV. Because of this low bandgap, Si can capture photons of wider range of the spectrum producing higher current. However the voltage generated is low due to the limitations of dark current loss. The indirect bandgap of Si causes relatively higher non radiative recombination loss. This occurs mostly due to the phonon emission at the time of band to band transition of electron. Moreover absorption coefficient of Si is lower than most of the III-V semiconductor materials. Thus reduced quantum efficiency further reduces the light to electricity conversion efficiency of a Si solar cell. Figure 1.1 depicts that only 33% of the incident photons contribute to the useful electricity generation in Si based single junction solar cells. Noticeably, most of the losses occur due to thermalization which deteriorates the cell performance further. Multijunction solar cell has emerged to overcome the losses incurred in single junction cells. III-V semiconductor materials are mostly used to build multijunction cells because of their excellent bandgap tuning capacity. These materials have direct electronic band gap which reduces non radiative recombination.

In multijunction approach the solar spectrum is effectively splitted by several junctions with each junction capturing a certain portion of the solar radiation spectrum. While commercially available silicon solar cell efficiency ranges from 8% to 18%, the existing multijunction cell can easily reach 37% conversion efficiency under 1 sun concentration. With this efficiency, electricity generation cost can be 28 ¢ /KWh [8]. With the ongoing research, it is

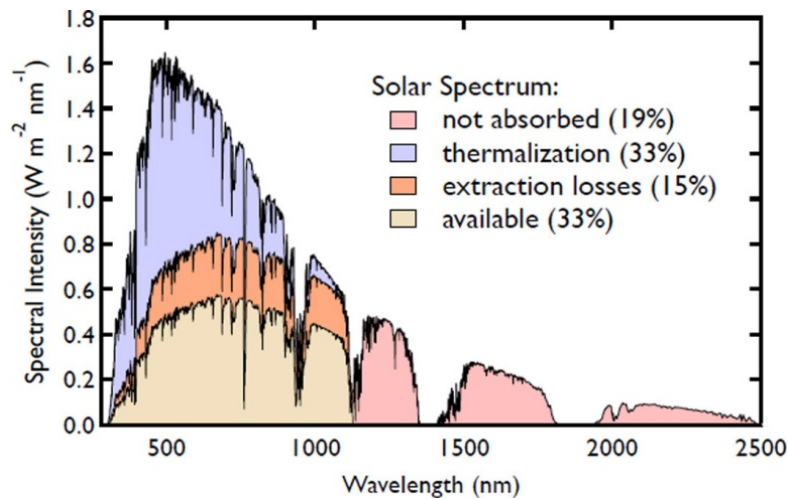

Fig. 1.1 Different losses in a silicon solar cell [7]

(Courtesy: SPIE-International Society for Optical Engineering)

projected that the cost of electricity generation from multijunction solar cells could exponentially decrease to as low as 7 ¢ /KWh by 2020 [9].

## 1.2 History of Multijunction Solar Cell Research

Multijunction solar cell research emerged with the goal to replace silicon solar cells with radiation resistant materials for space applications. GaAs showed excellent radiation hardness which guided the research community to develop GaAs devices on Ge substrate in late 1970s [10]. Bedair et al. reported the first epitaxially grown double junction monolithic solar cell based on AlGaAs/GaAs in 1979 [11]. Olsen et al. developed a similar solar cell based on GaInP/GaAs layers, which was utilized for space missions in mid 1990s [12]. In the year 2000, the first triple junction cell based on GaInP/GaInAs/Ge was grown on Ge substrate [10]. It was extensively used in spacecrafts during that time. Multijunction cells offer higher efficiency than any other type of solar cells e.g. Si, GaAs, CdTe or perovskite based single junction cells. The solar cell efficiency chart in figure 1.2 clearly reflects this fact. Noticeably, high concentration photovoltaics (HCPV) grew in parallel with the multijunction approach. Multijunction technology in fact acted as a driving force for III-V material based HCPV. According to the chart, Japan Energy achieved 31.4% single sun efficiency in 1996 from a dual-junction cell structure. The research collaboration between the Boeing Spectrolab and the National Renewable Energy Laboratory (NREL) achieved 32.3% concentrated efficiency (47 sun concentration) from lattice matched GaInP/GaInAs/Ge cell in 2000 [13]. Boeing Spectrolab achieved a milestone in the year 2006 by demonstrating 40.7% of power conversion efficiency in a metamorphic GaInP/InGaAs/Ge cell under 240 sun concentration [14]. The next year, NREL reported 34.1% single sun efficiency for GaInP/GaAs/InGaAs triple-junction solar cell with inverted metamorphic structure [15]. In 2009 Sharp surpassed this record by achieving 35.8% efficiency with the same material having better quality and structure [16]. By that time Fraunhofer Institute and Boeing Spectrolab had developed some other inverted metamorphic triple junction solar cells, which produced efficiency in between 41-44% under high sun concentration. In 2013, after further refinement of their triple junction inverted metamorphic technology, Sharp attained 37.9% efficiency which is the present day record efficiency for triple junction cell under single sun illumination. The attempt of increasing number of junctions is also noticed. Spectrolab devised a five-junction cell by wafer bonding as a triple junction cell grown on GaAs substrate and a dual junction cell developed on InP substrate [17]. It produced 38.8% efficiency under single sun condition.

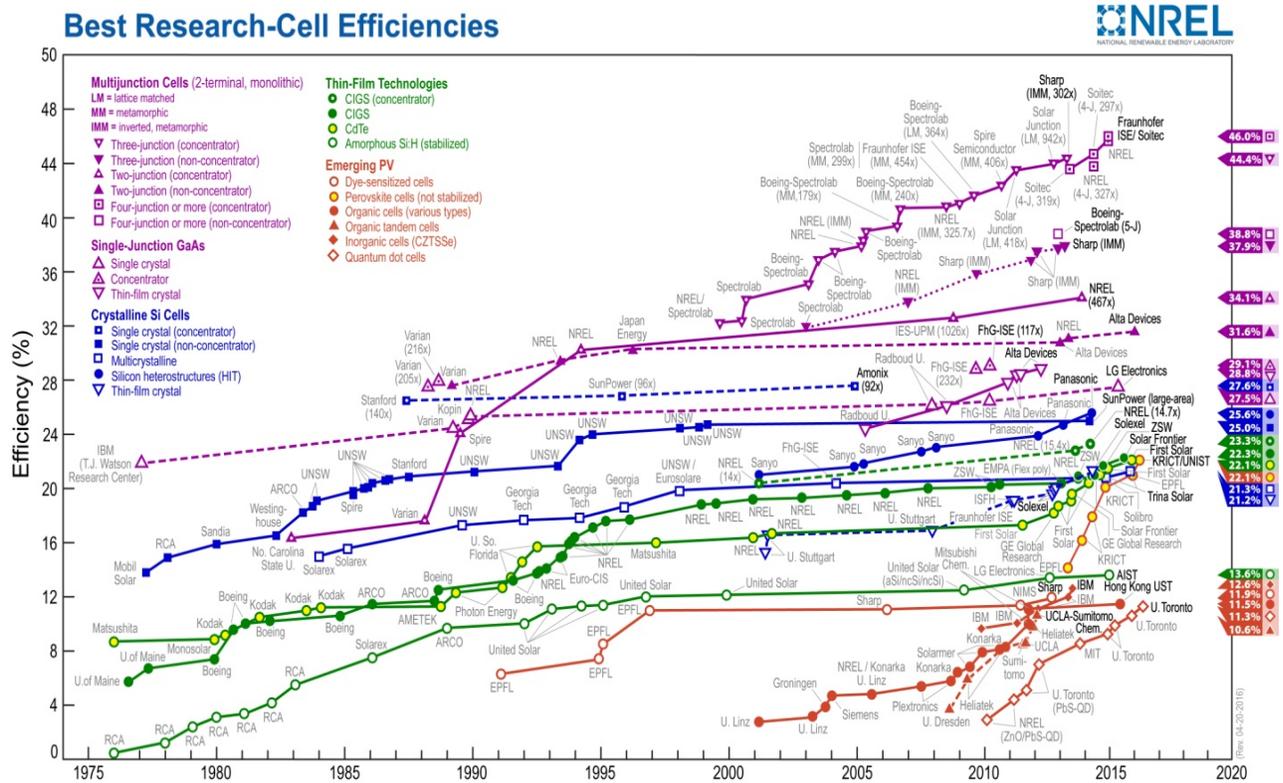

Fig. 1.2 Solar cell efficiency chart [18] (Courtesy: NREL-National Renewable Energy Lab)

When conversion efficiency did not increase further for unconcentrated structure, it increased considerably for concentrated structure, especially with higher number of junctions. The record efficiency of 46% was reported by Fraunhofer Institute in 2014 for a quadruple junction solar cell under 297 sun concentration [19].

### 1.3 Motivation behind the Research

*"And leave we will.*
*Yet as long as I breathe,*
*I will go on clearing the debris*
*with all my strength*
*From the face of this earth.*
*I will make this world habitable for this child;*
*This is my firm pledge to the newborn"* ---------- *Sukanta Bhattacharya [1926-1947]*

To me, acquiring knowledge becomes much more meaningful when it is utilized to improve the society. There is no doubt that generating usable form of energy to meet the growing demands of the civilization is a challenge for the scientists and engineers all over the world. Most of the countries in the world are not capable of generating power enough to support their economy and to ensure improved lifestyle for their citizens. I came from typical village in Bangladesh, where until today there is no electricity. I can remember studying with a kerosene lamp in my childhood days. Like other under developed countries, Bangladesh cannot import necessary amount of fossil fuels and build large infrastructures for power plants with its small economy. However, countless number of photons are falling on earth every day. We do not have to pay for it; we just need to build a suitable infrastructure to utilize this abundant energy. Unlike fossil fuels, sunlight is pollution free which reinforces our pledge of building a smarter and greener planet. Considering these facts I decided to work on solar cells. One important reason for solar cells not being popular is their low efficiency and high installation cost. Most of the commercially available solar cells are single junction. The low efficiency of these cells comes from their inability to capture all the photons coming from sun and several loss mechanisms. Multijunction solar cells in contrast utilize most of the photons incident on it. Bandgap engineered III-V compounds can be used to efficiently distribute solar spectrum among the junctions depending on their bandgaps. The present record efficiency InGaP(1.83 eV)/GaAs(1.40 eV)/InGaAs(1 eV) triple junction solar cell utilizes photons photons only in the range of 677 nm-1240 nm leaving the ultraviolet and most of the infrared photons unused. Solar spectrum ranges from 280 nm to 2500 nm. Thus using a higher bandgap material at the top and adding another junction with low bandgap material can increase the solar conversion efficiency. This was the biggest motivation of this research. Ge which has bandgap of 0.66 eV is not a good candidate as bottom subcell layer because it is an indirect bandgap semiconductor and it can capture only up to 1878 nm. Finding an even lower bandgap, III-V direct bandgap material with relatively less lattice mismatch with the InGaAs, was one of the important goals of this research. In this thesis, my primary goal was to design a cost-effective multijunction solar cell that will provide very high efficiency. Higher efficiency will surely reduce the cost per unit electricity production. I hope, clean electricity generated from photovoltaics will one day be very affordable and find its application in each and every sector of the human civilization. If coal fired power plants are shut down, vehicles on the road run with the electricity produced from the PV systems mounted on their surface, rivers do not die due to hydroelectricity production using dams, we will get a pollution free earth. I am happy that I have been able to do 'something' towards building a better habitat for the coming generations.

## 1.4 Organization of the Thesis

Chapter-2 is the study of the theoretical aspects behind the operation of a solar cell. At first the nature of light and dependence of light-matter interaction mechanism on wavelength are explained. Then some fundamental semiconductor properties such as crystal structure and electronic band structure, PV cell electrostatistics, electron-hole pair generation mechanism, radiative and non-radiative recombination methods and minority carrier lifetime is discussed. The structure and electrical model of a single junction solar cell are also illustrated to give a clear idea about solar cell operation.

The key aspects of multijuction solar cell technology are introduced and thoroughly explained in Chapter-3. Beginning with the operating principle, different possible configurations and the challenges associated with their realization, device performance based on the level of illumination, design factors and absorption characteristics of the materials are discussed. Different epitaxial methods used for growing a monolithic multijunction cell are also explained. After addressing different options and issues related to the design and development, different key considerations to design high efficiency tandem solar cells are also suggested.

The proposed quadruple junction solar cell is presented in Chapter-4, following literature review and the analysis of the design aspects. The structure is illustrated and the material properties assumed for the design is discussed in details. As part of the design methodology, the quantum efficiency and current density of all the subcells of the proposed design and their dependence on thickness and doping level are explained. The illustration of different electrical parameters such as short circuit current, open circuit voltage a fill factor associated with the J-V curve is mentioned. Different optimization attempts were undertaken to achieve the highest efficiency possible by changing doping level and thicknesses. Finally, the impact of diode non-ideality on device performance was investigated.

Chapter-5 presents some realistic analysis of the proposed quadruple junction solar cell for space applications. Modifications in design undertaken to achieve current matching and higher efficiency are mentioned. Last but not the least, the future scope of the work has also been discussed.

# CHAPTER 2

# THE PHYSICS OF SOLAR CELLS

## 2.1 Introduction

Solar cells are semiconductor devices capable of absorbing light and transforming it into electricity. This transformation happens though a complex light matter interaction involving absorption, electron-hole pair generation, recombination and carrier transport mechanism. Before designing a solar cell, it is essential to understand the working principle of photovoltaic devices. In this chapter at first we will explore the nature of light. Then the semiconductor physics with some important terms will be discussed. Finally we will explore the mechanism through which electron-hole pairs are generated and collected in a solar cell.

## 2.2 Sunlight

### 2.2.1 Nature of light

What is light? This is a fundamental question scientists and philosophers have thought about over years. It is well established that light is a special kind of electromagnetic energy. Scientists have observed that light energy can behave like a wave as it moves through space, or it can behave like a discrete particle with a discrete amount of energy (quantum) that can be absorbed and emitted. Both of these models are helpful in understanding and explaining the physical phenomena related to light.

Light is an electromagnetic wave having wavelengths between gamma rays and radio waves. The wavelength of solar radiation ranges from 200 nm to 2500 nm. We can see a small portion of this radiation (from 400 nm to 750 nm) as depicted in Fig 2.1. When we see light, the sense of colors in our brain depends on the wavelength of the light fallen on our eye. Blue light has the shortest and red light has the longest wavelength among the colors; green, orange and yellow being situated in between.

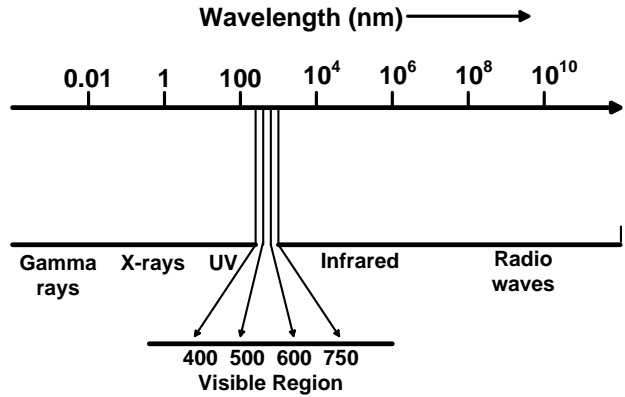

Fig. 2.1 The relative position of light in the electromagnetic spectrum

In the particle model, light consists of discrete energy particles called photons. A photon has no mass and charge. It carries a packet of electromagnetic energy when it travels from one place to another and interacts with other discrete particles (e.g. electrons, atoms and molecules) when it falls on or passes through a medium. The dual nature of light allow us to calculate the energy of a photon as a function of wavelength, $\lambda$. The photon energy $E_\lambda$ can be expressed as,

$$E_\lambda = \frac{hc}{\lambda} \qquad (2.1)$$

Here, $E_\lambda$ is in joules, $\lambda$ is in meters, $h$ is the Planck's constant (6.625 x 10$^{-34}$ Js) and $c$ is the speed of light in vacuum (2.998 x 10$^8$ ms$^{-1}$). From equation 2.1, we can easily calculate that the blue light consists of the most energetic and red light the least energetic photons. Only photons with energy greater than or equal to a semiconductor bandgap contribute to the photo-electricity generation.

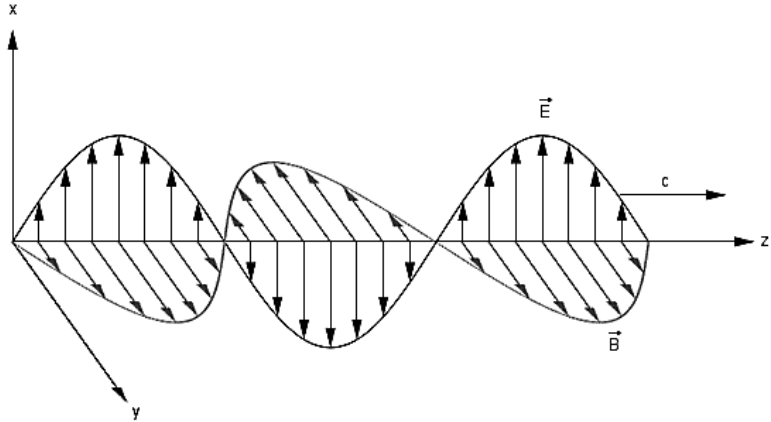

Fig. 2.2 Light as an electromagnetic wave [1]

Like other electromagnetic waves, light is composed of mutually perpendicular electric field E and magnetic field B, the direction of propagation being perpendicular to the both. The amount of energy a wave carries across a unit perpendicular area in every second is called irradiance of light.

The wave model of light is important to explain some natural phenomena like polarization, reflection, refraction, diffraction, superposition, interference etc.

### 2.2.2  Solar Spectrum

Sun as a star is a perfect sphere full with hot plasma [2], is the most important source of energy for life on earth. This radiation is isotropic. However due to earth's great distance from sun (approximately 150 million kilometers), only the photons emitted directly towards the earth reaches the earth's surface. Therefore, sunlight on earth can be considered as parallel stream of photons. With the temperature of 5777 K on Sun's outer surface [3], solar radiation can be considered as a blackbody radiation. Power density $W_s$ radiated from a unit area of a black body of temperature $T$ can be found from Stefan-Boltzmann law [4]:

$$W_s = \sigma_s T^4 \qquad (2.2)$$

Here $W_s$ is in watt/m², $T$ is temperature in Kelvin and $\sigma_s$ is the Stefan Boltzmann constant (5.67 x $10^{-8}$ watt/m² * $K^4$). Putting $T$ =5777 K in equation 2.2, we see that an enormous power of 63.15 MW flows from just 1m² area in sun's surface. This radiation gets attenuated to a great extent after travelling through a huge distance. When it reaches the earth's atmosphere, it falls to 1353 W/m² [5]. When sunlight enters into the earth's atmosphere, it is further attenuated due to scattering, reflection and absorption in the earth's atmosphere. If the light beams fall perpendicular to the earth's surface, the average irradiance becomes 1000 W/m² after experiencing minimum losses due to the absorption, according to American Society for Testing and Materials (ASTM). Out of this 1,000 W/m², 30 W/m² is of ultraviolet radiation, 440 W/m² is of visible light and the rest is infrared radiation. However scattering and reflection loss occurs if it falls on earth's surface in an inclined path. In this case power density is lower than 1000 W/m². The power density at earth's surface, depends on Air Mass (AM). It is in fact the depth of the atmosphere the solar radiation has to travel before falling on earth's surface. The air mass number can be expressed as [6],

$$AM = \frac{1}{\cos(\theta)} \qquad (2.3)$$

Here θ is the zenith angle of sun's position, perpendicular to earth's surface. Any radiation that has not reached earth's atmosphere has the number AM0, since it has travelled no distance in atmosphere. When the sun is directly overhead, air mass becomes 1 on earth's surface. Similarly a 48.2° inclination angle makes the sun spectrum an AM1.5 radiation. AM1.5 spectrum normalized to a total power density of 1000 W/m² is generally used as a standard for comparison of solar cell performance. In earth's surface there is an indirect diffuse component of spectral content due to the scattering and reflection phenomena in the atmosphere and rough landscape. Therefore, AM1.5 solar spectra can be further classified into global AM1.5 and direct AM1.5 respectively.

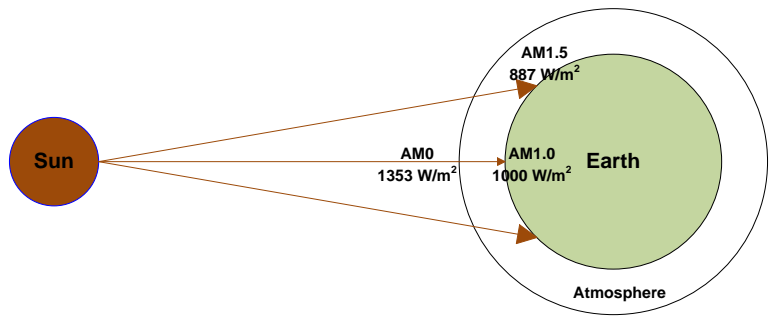

Fig. 2.3 Difference of solar intensities at three distinct points

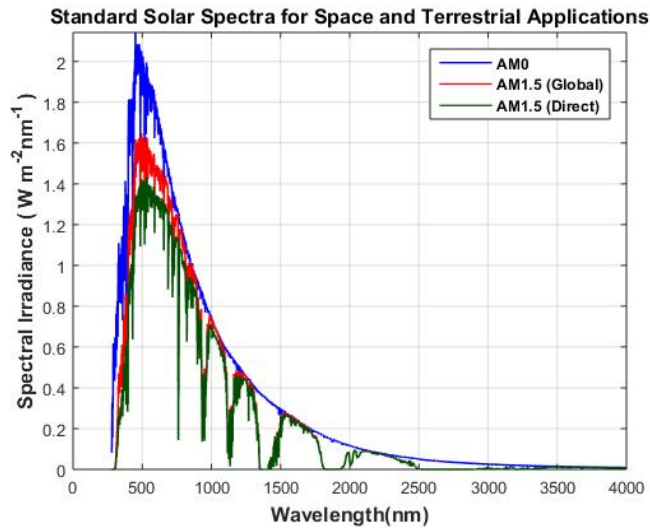

Fig. 2.4 Spectral irradiance vs wavelength

The diffuse component is considered in AM1.5G (i.e. global) but not in AM1.5D (i.e. direct). AM0, AM1.5G and AM1.5D are the standard solar spectrums for the space, terrestrial and concentrated photovoltaic efficiency calculation respectively. The corresponding power densities are 1356 W/m$^2$, 1000 W/m$^2$ and 887 W/m$^2$ respectively [7]. The difference in solar intensities in these three distinct points in space is depicted in Fig. 2.3. However, power density also varies with different wavelength values. The power density (power per unit area) of a particular wavelength is called spectral irradiance. Since wavelength values are normally expressed in nm, the unit of spectral irradiance is W m$^{-2}$ nm$^{-1}$. The variation of spectral irradiance is evident in the Fig. 2.4, which has been simulated using the data found from the American Society for Testing and Materials [8]. As illustrated in the figure, sun light ranges from ultraviolet to infrared wavelengths (280 nm to 2500 nm), having the maximum radiation intensity in the visible range. More specifically, maximum intensity is in the green portion of the solar spectrum having the wavelength around 500 nm. Using equation 2.1, we can conclude that the energy of photons ranges from 0.496 eV to 4.43 eV. The difference in spectral densities for all the standards is also noticeable. As expected AM0 has the highest value of spectral density for any wavelength. AM1.5D, representing only the photons that hit earth surface directly, on the other hand has the lowest spectral density.

Photon flux is an important parameter for solar cell performance. It is the number of photons falling on unit area in unit time. If $SI_\lambda$ is the spectral irradiance in W m$^{-2}$ nm$^{-1}$ and $E_\lambda$ is the energy of a single photon in joules, then the photon flux $\phi$ can be expressed as following:

$$\phi = \frac{\sum_{\lambda=280nm}^{\lambda=2500nm} SI_\lambda \times \lambda}{E_\lambda} = \frac{\sum_{\lambda=280nm}^{\lambda=2500nm} SI_\lambda \times \lambda}{\frac{hc}{\lambda}} \qquad (2.4)$$

Here, the expression for $E_\lambda$ has been placed based on equation 2.1. The importance of the photon flux depends on the bandgap of a semiconductor material used. Only those photons having energy higher than the bandgap of a semiconductor used contribute in solar energy generation.

## 2.3 Fundamental Semiconductor Properties for Solar Cells

Solar cells are generally fabricated from semiconductor materials for example silicon (Si), germanium (Ge), Gallium Arsenide (GaAs), Gallium Indium Phosphide (GaInP), Copper Indium Di Selenide (Cu(InGa)Se$_2$), Cadmium Telluride (CdTe) etc. Resistivity of a semiconductor material ($10^{-4}$ - 0.5 $\Omega$m) [9] lies in-between conductors and insulators. This resistivity value makes it suitable to control the flow of electrons and holes in these materials. Also, they show a negative temperature coefficient of resistance i.e. resistance decreases with increase in temperature and vice-versa. However resistivity can be easily changed by adding a small amount of trivalent or pentavalent impurity to these materials. This process is called doping. The new materials after doping with trivalent and pentavalent material are called p type and n type semiconductors respectively. Holes and electrons are the majority carriers in these materials respectively. When a p and an n type material are joined together they form a p-n junction. The insulating region formed in their interface is called a depletion region or a space charge region. A single junction solar cell is nothing but a p-n junction where carriers are generated using the energy of sunlight.

Semiconductor materials can be just elemental like Si, Ge etc or compound in nature. When two semiconductor elements are combined, it is called a binary compound. These compounds are formed by adding a group II semiconductor with a group VI semiconductor like

CdTe, ZnSe etc. It is also possible to form a binary compound by adding a group III semiconductor with a group V semiconductor. GaAs and GaP are two good examples of such a combination. When three semiconductor materials are combined together to form a compound, it is called a ternary compound. Some good examples are $Al_xGa_{1-x}As$, $Ga_xIn_{1-x}P$ and $Ga_xIn_{1-x}As$, where x stands for the fraction of composition. By changing the value of x, the electrical and optical properties of a semiconductor compound can be easily varied. This is an effective technique for finding out the right material composition and meeting a certain value of electronic or optical criterion (e.g. electronic bandgap, lattice constant, phonon bandgap etc.) for a particular application.

To understand the operation of a solar cell properly, we have to become familiar with some basic concepts of solid state physics.

### 2.3.1 Crystal Structure

Most of the semiconductors are crystalline in nature i.e. their atoms are aligned in a regular periodic pattern. The group of atoms which create such a pattern is called a basis. A crystal structure is formed when several basis units are repeated periodically. The crystal structure along with the atomic properties of the material determines the electrical behavior of the material. The most common crystal structures are body centered cubic (BCC), face centered cubic (FCC) and hexagonal closed pack (HPC) crystal lattice. There is another simple structure named as simple cubic (SC) which is an idea and used for didactical purpose only. There is no such material in reality which has perfect SC structure.

In a SC structure atoms lie in the eight corners of a cube. FCC structure is an extension to the SC structure where one atom lies in the middle of the cube. In FCC structure, six atoms lie in the middle of the six faces of the cube, in addition to simple cube. In the HCP structure, layers of atoms are packed so that atoms in alternating layers overlie one another. Many semiconductors have either diamond or zincblende structure. For example, silicon is a group IV element which has four electrons in its outer shell. Atoms in crystalline silicon are arranged in diamond lattice by tetrahedral bonding. On the other hand many group III-IV and II-VI semiconductors have zincblende structure. As an example, GaAs has zincblende structure which is formed by two interpenetrating FCC unit cells; one entirely composed of gallium another of arsenic. All the four materials used in the design of the novel quadruple junction solar cell have the same zincblende structure.

### 2.3.2 Electronic Band Structure

An electron moving through a semiconductor material is analogous to a particle in a three dimensional box because of the tightly bound electrons in the inner shells and the potential fields from the surrounding atoms' nuclei. Schrödinger equation is an important tool for describing the dynamic behavior of a particle in a quantum mechanical box. The time-independent one-dimensional Schrödinger equation can be expressed as [10],

$$\nabla^2 \psi(x) + \frac{2m}{\hbar^2}[E - U(\vec{r})]\psi(x) = 0 \qquad (2.5)$$

Here, $\psi$ is the time independent wave function, m is the mass of an electron (9.11 x $10^{-31}$ Kg), $E$ is the energy of the electron and $U(\vec{r})$ is the potential energy of the electron inside the semiconductor. Also, $\hbar$ is the reduced Planck's constant i.e. $\hbar = \frac{h}{2\pi}$, where h=6.626 x $10^{-34}$

$m^2kgs^{-1}$. Solving this differential equation we get,

$$\psi(x) = Ae^{jkx} + Be^{-jkx} \qquad (2.6)$$

Here A and B are two constants. $k$ is the wave number. This solution suggests that wave propagates in both the positive and negative direction. That means, the quantum mechanically computed motion of electron in the crystal is like that of an electron in free space if we replace its mass m by the effective mass $m^*$. From classical mechanics we know,

$$F = m^* a \qquad (2.7)$$

Here $a$ is the acceleration of the electron when force $F$ is applied on it. Now the relation between energy $E$ and momentum $p$ can be written as,

$$E = \frac{p^2}{m} = \frac{k^2 \hbar^2}{2m} \qquad (2.8)$$

From this equation it can be noticed that the value of energy is a discrete number. According to Pauli's exclusion principle, the total energy in the crystal lattice falls into different bands of energies. However they are separated into two groups; one is called conduction band and another is called valence band. The number of energy levels in each of these two groups is expressed by the density of the states. Density of states depends on the property of the semiconductor material and the temperature of the environment. The square term in the right side of equation 2.8 suggests that the energy-momentum diagram is parabolic in nature.

The electrons lying in the valence band are those orbiting round the nucleus from the outermost shell. On the other hand electrons in the conduction band are free to move. There is a barrier between these two bands which is called forbidden energy gap or simply bandgap of the material. Bandgap $E_g$ can be expressed as,

$$E_g = E_c - E_v \qquad (2.9)$$

Here $E_c$ is the lowest value of energy in conduction band and $E_v$ is the highest value of energy in the valence band. An electron in the valence band has to attain at least $E_g$ amount of energy to jump into the conduction band and start flowing through the lattice. This energy can be provided from an external source like light or heat. The valence to conduction band transition can happen in two different ways. Based on the transition process semiconductor materials are classified into two types: direct bandgap and indirect bandgap semiconductor. $N_A$

### 2.3.3 Direct and Indirect Bandgap Semiconductor

In a direct bandgap semiconductor, the peak of the valence band and the valley of the conduction band are situated in the same value of momentum. Therefore, when photons of energy $E_g$ or greater falls on such a material, electrons can easily jump to the conduction band (i.e. starts conducting through the lattice) leaving a hole in the valence band. Thus it is an easy process once sufficient amount of energy is provided. InAs, GaAs, GaP, CdTe and $Cu(InGa)Se_2$ are some direct bandgap semiconductors. There are some other semiconductor materials where the peak of the valence band and the valley of the conduction band do not lie in the same level of electron momentum. Therefore an electron cannot directly jump to the conduction band; it has to attain the same momentum as in the valley of conduction band at first. This is done by phonon (vibration in the lattice) absorption or emission as depicted in figure 2.5b. This is not an easy process as compared to the direct band transition approach illustrated in figure 2.5a.

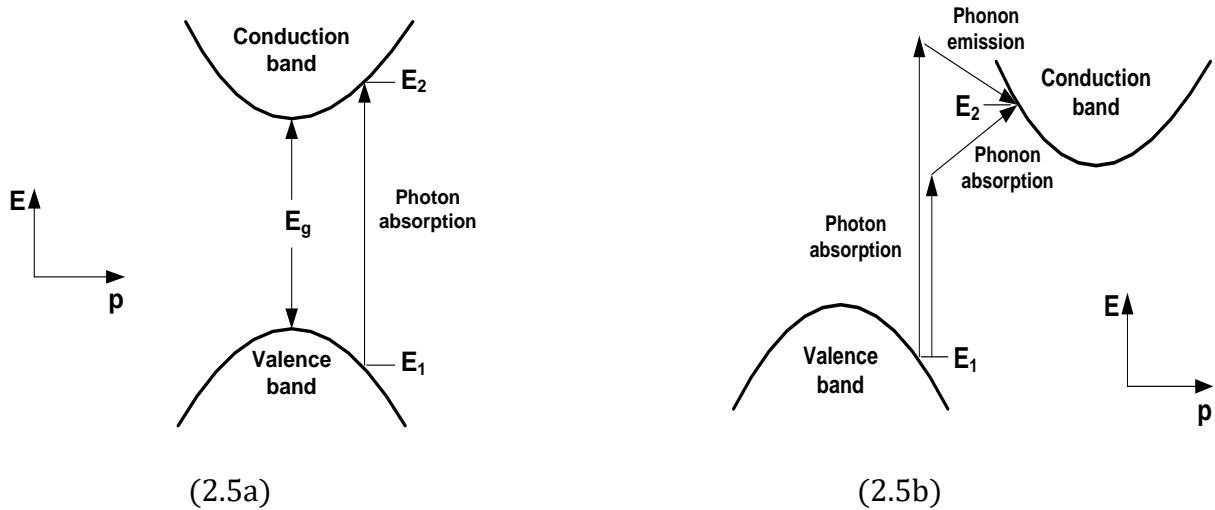

(2.5a)                  (2.5b)

Fig. 2.5 Valence-Conduction band transition approach in (a) direct and (b) indirect bandgap semiconductor [6]

Since both a phonon and electron are needed, the absorption coefficient depends not only on the full initial electron states and empty final electron states, but also on the availability of the absorption or emission phonon with the required value of momentum. Hence the number of electron transition in the indirect bandgap material is lesser than that in the direct bandgap material. As a result photon can penetrate an indirect bandgap material more without being absorbed. However many well known materials like Si and Ge are indirect bandgap semiconductors.

    Bandgap engineering allows us to design a material as direct or indirect bandgap semiconductors. However in most of the cases a direct bandgap semiconductor is preferred for its simplicity in band to band transition approach.

### 2.3.4  PV cell Electrostatics

A photovoltaic cell is nothing but a *pn*-junction formed by placing a *p*-type material in contact with an *n*-type material. A depletion region (in other words, quasi neutral region) is formed around the contact surface. The electrostatic potential arising from the junction formation is called built in potential, which is denoted by $V_{bi}$. The electrostatics of a *pn* junction diode can be expressed by Poisson's equation,

$$\nabla^2 \phi = \frac{q}{\varepsilon}(n_0 - p_0 + N_A^- - N_D^+) \qquad (2.10)$$

Here $\phi$ is the electrostatic potential, $q$ is the charge of an electron, $\epsilon$ is the permittivity of the semiconductor, $n_0$ is the electron concentration and $p_0$ is the hole concentration in equilibrium condition. Also, $N_D^+$ is the ionized donor concentration and $N_A^-$ is the ionized acceptor concentration respectively. With reference to the figure 2.6, the depletion region can be defined as $-X_N < X < X_P$. We can assume that within the depletion region, $p_0$ and $n_0$ both are negligible in comparison to $|N_A - N_D|$. Equation 2.10 can be written in simplified form as,

$$\nabla^2 \phi = -\frac{q}{\varepsilon} N_D \text{ for } -X_N < X < 0 \qquad (2.11a)$$

$$\nabla^2 \phi = \frac{q}{\varepsilon} N_A \text{ for } 0 < X < X_P \qquad (2.11b)$$

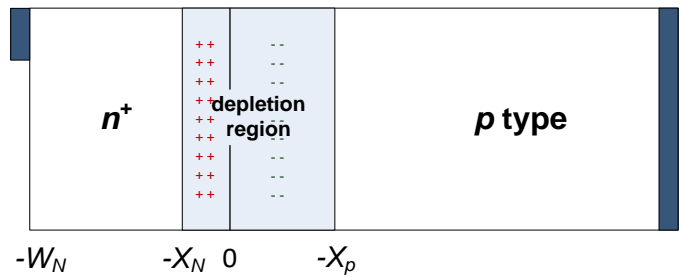

Fig. 2.6 Illustration of simple solar cell operation

Outside the depletion region, charge neutrality is assumed. Now we can obtain the built-in voltage, $V_{bi}$ by integrating the electric field. We know that,

$$\vec{E} = -\nabla \phi \qquad (2.12)$$

Therefore, $\int_{-x_N}^{x_P} \vec{E} dx = -\int_{-x_N}^{x_P} \frac{d\phi}{dx} dx = -\int_{V(-x_N)}^{V(x_P)} d\phi = \phi(-x_N) - \phi(-x_P) = V_{bi} \qquad (2.13)$

Solving equation 2.11 and using $\phi(x_P) = 0$, we get

$$\phi(x) = \begin{cases} V_{bi} & x \leq -x_N \\ V_{bi} - \frac{qN_D}{2\varepsilon}(x+x_N)^2 & -x_N < x \leq 0 \\ \frac{qN_D}{2\varepsilon}(x-x_P)^2 & 0 \leq x < x_P \\ 0 & x \geq x_P \end{cases} \qquad (2.14)$$

Now, from Einstein's relationship in 1D, the electric field can be written as,

$$\vec{E} = \frac{kT}{q} \frac{1}{p_0} \frac{dp_0}{dx} \qquad (2.15)$$

Utilizing this in equation 2.13, we get

$$\int_{-x_N}^{x_P} \vec{E} dx = \int_{-x_N}^{x_P} \frac{kT}{q} \frac{1}{p_0} \frac{dp_0}{dx} = \frac{kT}{q} \int_{p_0(-x_N)}^{p_0(x_P)} \frac{dp_0}{p_0} = \frac{kT}{q} \ln\left[\frac{p_0(x_P)}{p_0(-x_N)}\right] = V_{bi} \qquad (2.16)$$

If we consider non-degeneracy, then $p_0(x_P) = N_A$ and $p_0(-x_N) = n_i^2/N_D$. So,

$$V_{bi} = \frac{kT}{q} \ln\left[\frac{N_D N_A}{n_i^2}\right] \qquad (2.17)$$

This built in potential plays a crucial role in solar cell operation. It amasses all the minority holes in p region and all the minority electrons in n region and thus contributes to the creation of photocurrent.

## 2.3.5 Electron-Hole Pair Generation Mechanism in Solar Cells

Metals cannot absorb and transform light into electricity. In case of insulators, although they can absorb, the generated electron energy is lost to a great extent due to the high resistance of the insulators. Semiconductors are the best choice in this regard. The absorption rate depends upon two things: absorption coefficient, $\alpha$ (expresses absorption capability) and the thickness of the material. If light of intensity $I_0$ falls on a solar cell top surface, the intensity of light at any distance $x$ is given by,

$$I = I_0 e^{-\alpha x} \quad (2.18)$$

When light passes through a medium, refraction occurs and some parts of the light get attenuated. This attenuation is generally taken into consideration by defining a complex refractive index, $\hat{n}$.

$$\hat{n} = n + ik \quad (2.19)$$

Here $n$ is the real part of the refractive index and $k$ is the imaginary part. The imaginary part is called as extinction coefficient of the medium. The absorption coefficient of a material, $\alpha$ is related to the extinction coefficient by following relationship:

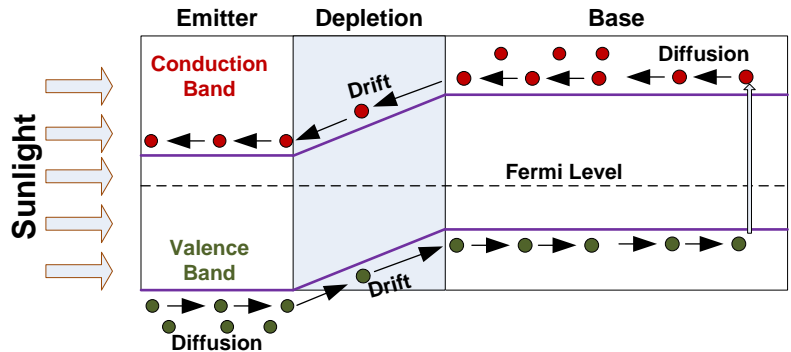

Fig. 2.7 Light absorption and carrier transport

$$\alpha = \frac{4\pi k}{\lambda} \qquad (2.20)$$

From this equation we can see that the absorption coefficient of a material varies with wavelength, λ. This suggests that a semiconductor material may be a good light absorber only for a particular wavelength range. As portrayed in figure 2.7, a photon of sufficiently high energy (i.e. sufficiently low wavelength) strikes an electron in the valence band and gets it excited to jump to the conduction band, leaving an empty hole in the valence band. Thus an electron-hole pair is created in base and emitter region respectively. The minority carriers in p type base and n type emitter diffuse towards the depletion region. When they enter into the depletion region, they are drifted due to the built in electric field, $V_{bi}$. The electrons created in the base region drift towards the emitter regions and holes created in the emitter regions drift towards the base region. This movement of minority carriers gives rise to a potential difference due to the splitting of Fermi level, $E_F$ into minority electron quasi Fermi level $E_{FN}$ and minority hole quasi Fermi level $E_{FP}$. This potential difference results in the open circuit voltage from the cell, $V_{oc}$ as,

$$V_{oc} = \frac{E_{FN} - E_{FP}}{q} \qquad (2.21)$$

Here, q is the charge of an electron (1.6 x $10^{-19}$ C).

### 2.3.6  Recombination

After light absorption the generated electrons and holes lie in conduction and valence band respectively. In the conduction band an electron stays at meta-stable state. If it is not collected timely, it goes back to the valence band and meets a hole there. When an electron meets a hole, both of them get annihilated. This process is called recombination.

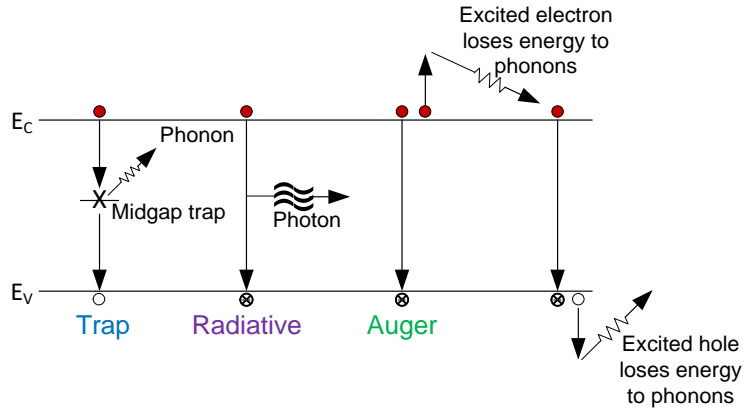

Fig. 2.8 Recombination methods in solar cells

The energy released from the annihilation process is either lost or utilized at a certain level later. Recombination methods can be classified into two groups mainly:

i. Radiative recombination method

ii. Non radiative recombination methods: Trap (Shockley-Read-Hall) recombination and Auger recombination

The recombination methods are illustrated in figure 2.8 and described below.

### 2.3.6.1 Radiative (band-to-band) Recombination

Radiative recombination is simply the reverse process of generation. In this case, the energy from electron-hole annihilation causes a photo emission. It is in fact the same process related to the operation of LASERs and LEDs. In case of spontaneous emissions (LED), the phase and direction are different than that of the incident light. In the case of stimulated emission (LASER), the emitted photons will have the same phase and direction as that of the incident light. The net radiative recombination rate can be written as [6],

$$R_{radiative} = B(pn - n_i^2) \quad (2.22)$$

Here B is a proportionality constant which is called radiative recombination coefficient; n and p are the concentrations of free electron and hole. Radiative recombination can easily be measured using the optical absorption spectrum of a semiconductor material. Group III-V semiconductors have lower radiative recombination compared to the other semiconductors like Silicon and Germanium.

### 2.3.6.2a  Trap (Schockley-Read-Hall) Recombination

It is not always possible to fabricate a solar cell defect free (it becomes cost ineffective). Trap recombination occurs due to the presence of impurity in the crystal. The recombination occurs in two steps. A free electron in the conduction band at first relaxes to the defect level and then relaxes to the valence band where it annihilates with hole in the valence band. This type is also called as Schockley-Read-Hall recombination. The energy given to an impurity defect creates phonon (lattice vibration) emission and finally gets lost as heat. The net trap recombination rate per unit volume per second through a single level trap is,

$$R_{trap} = \frac{pn - n_i^2}{\tau_{p0}(n + n_1) + \tau_{n0}(p + p_1)} \quad (2.23)$$

Here $\tau_{p0}$ and $\tau_{n0}$ are the fundamental hole and electron lifetimes. The general expression of fundamental carrier (either electron or hole) lifetime is,

$$\tau_0 = \frac{1}{\sigma V_{th} N_T} \quad (2.24)$$

Thus fundamental hole and electron lifetimes depend on the capture cross section $\sigma$ ($\sigma_n$ for electrons and $\sigma_p$ for holes), the thermal velocity of the carriers $V_{th}$ and the concentration of traps $N_T$. The factors $n_1$ and $p_1$ are statistical entities which can be determined from trap energy level $E_T$ as shown in equation 2.25a and 2.25b.

$$n_1 = N_C e^{\frac{E_T - E_C}{KT}} \quad (2.25a)$$

$$p_1 = N_V e^{\frac{E_V - E_T}{KT}} \quad (2.25b)$$

Here K is the Boltzmann's constant (1.38 x $10^{-23}$ $m^2 Kgs^{-2} K^{-1}$) and T is the temperature. $N_C$ and $N_V$ are the effective density of states in the conduction and valence band respectively.

### 2.3.6.2b  Auger Recombination

Auger recombination is similar to the radiative recombination, except that the energy of transition is given to another carrier in either the conduction band or in the valence band, as illustrated in figure 2.8. This electron (or hole) then relaxes thermally (both energy and momentum) by generating phonons. The net recombination rate due to Auger process is [11],

$$R_{Auger} = (C_n n + C_p p)(pn - n_i^2) \quad (2.26)$$

Here $C_n$ and $C_p$ are called Auger coefficients. Auger recombination affects solar cell performance mostly in case of high carrier concentrations caused by heavy doping or high level injection under concentrated sunlight. In silicon-based solar cells (most of the commercially available cells), Auger recombination limits the lifetime and ultimate efficiency. The more heavily doped the material is, the shorter is the Auger recombination lifetime.

### 2.3.7 Minority Carrier Lifetime

If minority electrons in conduction band (in excited state) are not collected within some time or if the minority carrier concentration rises very high with respect to the bulk volume, the excess minority electrons will recombine with majority holes. The time between generation and recombination is termed as minority carrier lifetime of a material. It is a crucial factor for solar cell performance. In low level injection materials (where the number of minority carriers is less than the doping) minority carrier lifetime is directly proportional to the excess minority carrier concentration Δn and inversely proportional to the recombination rate R as given in equation 2.27.

$$\tau = \frac{\Delta n}{R} \qquad (2.27)$$

However, recombination may occur both in surface and bulk of the semiconductor. Therefore, there are two lifetimes, $\tau_s$ and $\tau_b$ associated with these two recombination methods. The effective minority carrier lifetime is the summation of both the surface and bulk recombination.

$$\frac{1}{\tau_{eff}} = \frac{1}{\tau_s} + \frac{1}{\tau_b} \qquad (2.28)$$

The recombination in the bulk may result due to radiative, trap or Auger process. Thus the minority carrier lifetime in the bulk region can be written as,

$$\frac{1}{\tau_b} = \frac{1}{\tau_{radiative}} + \frac{1}{\tau_{trap}} + \frac{1}{\tau_{Auger}} \qquad (2.29)$$

## 2.4 Solar Cell Fundamentals

### 2.4.1 Structure

A solar cell is made by sandwiching an *n-type* emitter with a *p-type* base layer. The emitter is made of higher doping but lesser thickness than base layer. This is done, because the mobility of minority electrons (in base) is higher than the mobility of holes (in emitter). Sunlight enters into the cell from the emitter side. An antireflection (AR) coating layer is used to minimize light reflection into the cell. Without this layer, much of the light would bounce off the surface of the cell. While designing an AR coating layer, the refractive index and the thickness of the material are the two most important things to be considered.

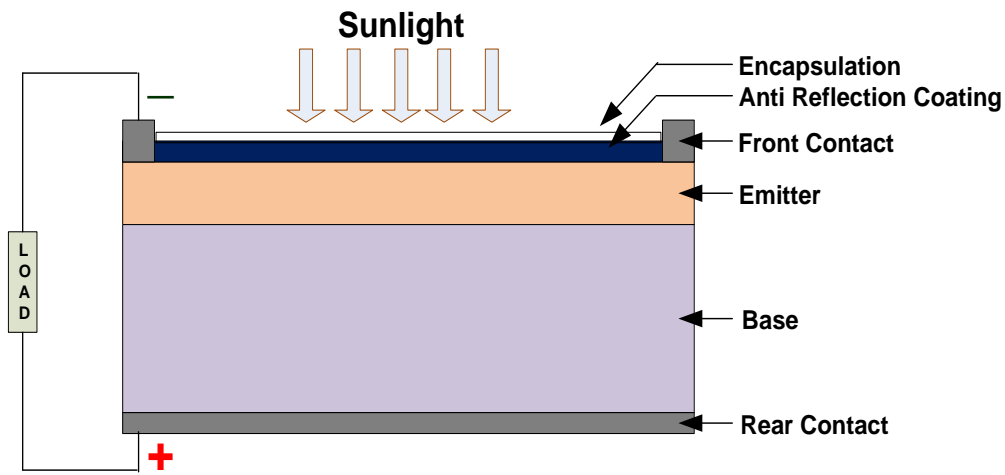

Fig. 2.9 Cross section of a solar cell

An encapsulation made of glass or plastic is placed on top of the AR coating. It protects the cell from the external environment. The front and back contacts are made of metals of good conductivity. They collect the photogenerated electrons and holes respectively. A cross section of a solar cell is given in figure 2.9. The rate of photogenerated carriers depends on the photon flux, energy of incident photons and the absorption capacity of the semiconductor. The absorption capacity depends on the performance of the antireflection coating, electronic bandgap of the semiconductor, intrinsic carrier concentration, carrier mobility, recombination rate, temperature and some other factors.

The solar cell structure described above is called a single junction solar cell, since there is only one junction formed in the interface between *n-type* emitter and *p-type* base. It is possible to form several junctions in a cell structure, each formed by a single set of emitter and base layers. This type of structure is called a multijunction cell (which will be discussed in the next chapter in details).

### 2.4.2    J-V Characteristics

#### 2.4.2.1    Solar Cell in Dark

Since solar cell converts light into electricity it may seem odd to analyze its characteristics in dark. However, it is an effective way to understand the electrical characteristics of a solar cell, because small fraction of light on illuminated cell can introduce noise in the operation. Let us imagine a solar cell in dark having the built-in voltage across the p-n junction as $V_{bi}$. Now, if an external voltage $V_A$ is applied in forward bias arrangement, the built in voltage will reduce to $(V_{bi} - V_A)$ and Fermi level $E_F$ will split to two quasi Fermi levels, $E_{FP}$ and E$_{FN}$. As the barrier width has reduced, electrons from n-side can easily diffuse to the p-region and holes from p-side can easily diffuse to the n-region. This process is called minority carrier injection and the current density created in the bulk (quasi neutral) region is called the recombination current J$_{rec}$. The recombination current is compensated by the thermal generation current J$_{gen}$ which is caused by the drift of minority carriers present in corresponding doped regions across the junction. When no voltage is applied to the p-n junction the situation inside the junction can be viewed as the balance between the recombination and generation currents i.e. J=J$_{rec}$- J$_{gen}$=0 for V$_A$=0. It is assumed that when moderate forward bias voltage is applied to the junction, the recombination current increases.

$$J_{rec}(V_A) = J_{rec}(V_A = 0)\left(e^{\frac{qV_A}{KT}} - 1\right) \quad (2.30)$$

The generation current is independent of the potential barrier across the junction and is determined by the availability of thermally generated minority carriers in the doped regions.

$$J_{gen} \approx J_{gen}(V_A = 0) \quad (2.31)$$

Thus the net current density can be represented as an exponential function of the applied voltage,

$$J = J_{rec}(V_A) - J_{gen}(V_A) = J_0\left(e^{\frac{qV_A}{KT}} - 1\right) \quad (2.32)$$

This equation is called Shockley's equation. Here J$_0$ is the saturation current density which depends on electron diffusion length (L$_n$), hole diffusion length (L$_p$), intrinsic carrier concentration ($n_i$) acceptor concentration ($N_A$), donor concentration ($N_D$) and diffusion constants ($D_n$ & $D_p$).

$$J_0 = qn_i^2 \left( \frac{D_n}{L_n N_A} + \frac{D_p}{L_p N_D} \right) \quad (2.33)$$

Hence saturation current density can be computed if the material properties are known.

### 2.4.2.2 Illuminated Solar Cell

When a solar cell comes under illumination, additional electron-hole pairs are generated which results in the increase of minority carrier flow. This flow of photogenerated carriers creates the so called photocurrent density $J_{ph}$ or $J_{pv}$ which adds to the thermal generated current $J_{gen}$.

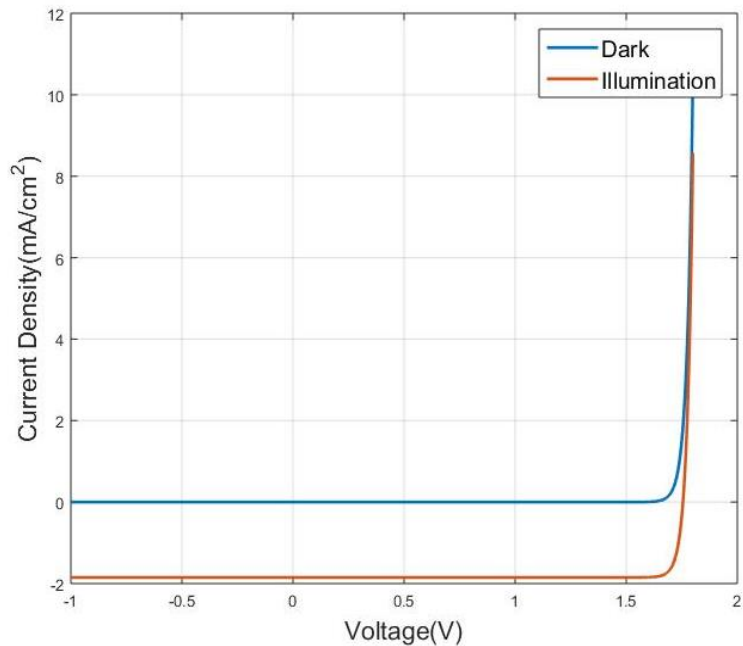

Fig. 2.10 Simulated J-V characteristics of a solar cell

Thus equation 2.32 can be written as,

$$J = J_{rec}(V_A) - J_{gen}(V_A) - J_{ph} = J_0 \left( e^{\frac{qV_A}{KT}} - 1 \right) - J_{ph} \qquad (2.34)$$

The equation above expresses that the illuminated J-V characteristics of the p-n junction is same as the dark J-V characteristic but it is shifted down by the photogenerated current density $J_{ph}$ as depicted in figure 2.10, which was simulated using equations 2.32 and 2.34. The photogenerated current density is a function of the minority carrier diffusion length L, width of the depletion region W and the uniform generation rate G.

$$J_{ph} = qG(L_n + W + L_p) \qquad (2.35)$$

It means that only carriers generated in the depletion region and the region comprising diffusion lengths contributed to the photogenerated current. One thing should be noted here, is that in solar cells there is no need of the biasing voltage. The actual current in fact is in the reverse direction similar to photodiodes. Therefore, it is customary to demonstrate J-V curve as a positive current-positive voltage curve i.e. the voltage is not the applied voltage, but the photogenerated voltage.

### 2.4.3 Electrical Model

A solar cell is similar to a *p-n* junction diode. Therefore, its characteristics equation is similar to a diode equation. For an ideal PV cell shown in figure 2.10, the basic semiconductor equation can be written as [15].

$$I = I_{pv} - I_{dark} = I_{pv} - I_0 e^{(\frac{qV}{aKT} - 1)} \qquad (2.36)$$

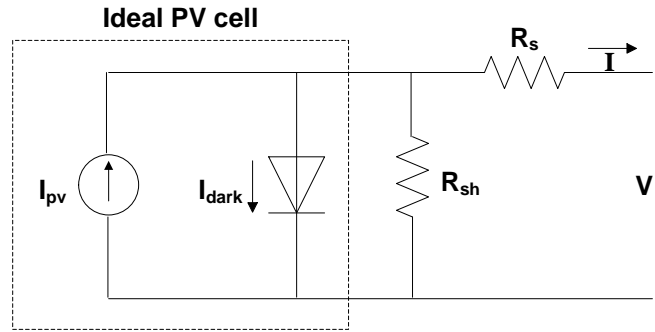

Fig. 2.11 Electrical model of a single junction solar cell

Here $I_{pv}$ is the photogenerated current. It is directly proportional to the solar irradiation. The amount of the resultant current I is limited by the dark current $I_{dark}$. The dark current can be the expressed as the current passing through a Shockley diode which has the reverse saturation/leakage current of $I_0$. Thus it depends on the photogenerated voltage V and temperature T. Here, q is the charge of an electron, K is the Boltzmann constant. and *a* is the diode ideality factor. For an ideal diode scenario, it equals to 1. However, practically it is always greater than 1. The ideal solar cell characteristics have been simulated with a diode ideality factor of 1.3. The electrical model is shown as figure 2.11. We can notice a tradeoff between the output current and the output voltage obtained from a solar cell. The reason behind is the choice of bandgap of the material. If higher bandgap material is chosen, there will be less number of electrons which will be able to overcome the forbidden energy gap, in the incidence of being energized by incident photons. In contrast, if a low bandgap material is chosen, a higher number of electrons will easily overcome the energy barrier obtaining energy from the incident photons. The power conversion efficiency of a solar cell is dependent upon the short circuit current (value of current when, V=0) and open circuit voltage (value of voltage when, I=0) in the I-V curve.

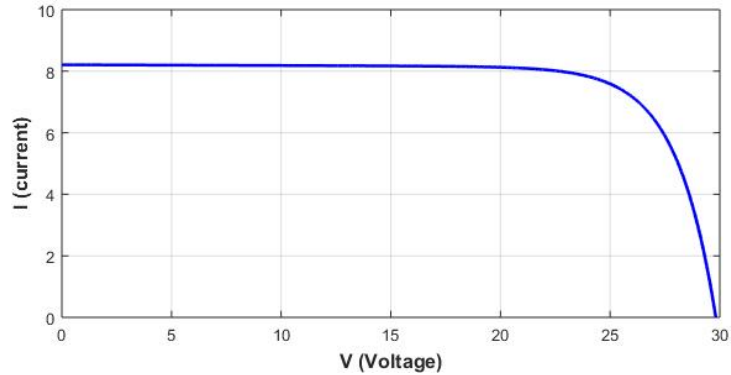

Fig. 2.12 Simulation of an ideal solar cell

A designer has to choose a bandgap value of the semiconductor so that maximum efficiency can be obtained.

In practical case, a solar cell does not have zero resistance; its total resistance can be modeled as a combination of series resistance $R_s$ and parallel resistance $R_p$ as depicted in figure 2.10. The source of these resistance values are the contact resistance and resistance in the bulk.

# CHAPTER 3

# MULTIJUNCTION SOLAR CELLS

## 3.1 Introduction

A multijunction solar cell consists of multiple p-n junctions made of different semiconductor materials. The use of multiple materials allows the cell to absorb broader range of solar spectrum and improves the solar conversion efficiency. Most of the commercially available solar cells are of low efficiency single junction cells. The theoretical efficiency limit of these cells is only 31% [1]. However the commercially available ones have efficiencies in between 8%-18%. This is undoubtedly not promising in meeting the growing energy demand of the world. With the maximum theoretical efficiency of 86.8% [2], multijunction solar cell technology has a huge potential. However, practical efficiency value of such a cell is limited by different losses. Due to the complexity in design and growth process, multijunction solar cells are costly and mostly used in space applications. Increase in efficiency and exponentially decaying fabrication cost will make multijunction technology cost effective for terrestrial applications also, considering its high throughput. A good design with proper semiconductor parameters considerations and suitable material usage may increase the conversion efficiency.

## 3.2 Why Multijunction solar cell?

The inability of single junction solar cells in absorbing the whole solar spectrum efficiently has led the researchers to multijunction approach. A single junction cell cannot absorb photons having energy less than the bandgap of the constituent material. The unutilized lower energy photons are wasted as thermalization loss. Only one electron-hole pair is generated if a photon of energy ($E_{ph}$) equal to or greater than the bandgap ($E_g$) falls on the solar cell. However, the excess energy ($E_{ph}$-$E_g$) results in lattice vibration (phonon emission), which later heats up the cell. In both cases, the thermalization of the lattice deteriorates the cell performance further. A multijunction cell uses multiple materials, each having different bandgaps. Therefore high energy photons are absorbed by higher bandgap materials and low energy photons by low bandgap material. Thus reducing the amount of the excess energy, it protects the cell from thermalization loss. As photon energy of all levels (ideally) is utilized to generate electricity, the light to electricity conversion efficiency is significantly higher than single junction solar cell.

### 3.3  Physics of Operation

A multijunction solar cell consists of several sub cell layers (or junctions), each of which is channeled to absorb and convert a certain portion of the sunlight into electricity. Each subcell layer works as a filter, capturing photons of certain energy and channel the lower energy photons to the next layers in the tandem. The light absorption in an *n-junction* solar cell is illustrated in figure 3.1. As depicted, the first junction absorbs only the ultraviolet photons having energy more than 2.75 eV [3] because it was made of semiconductor material having bandgap of 2.75 eV. Since violet, blue, green, yellow, orange, red and infrared photons have energy lower than 2.75 eV, they cannot provide enough energy for an electron to jump to conduction band. Thus they cannot be absorbed by the first layer and are transferred to the next layer.

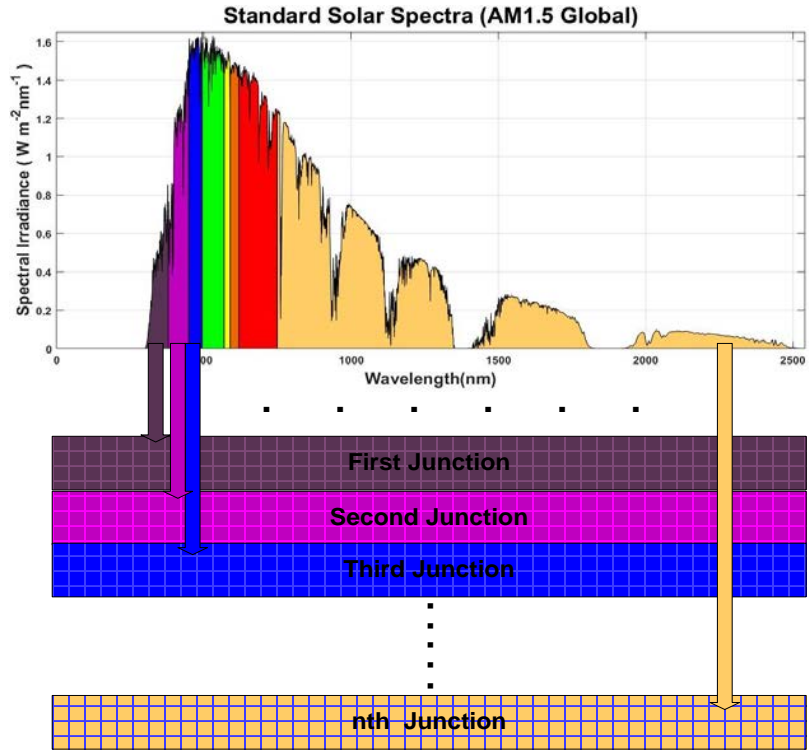

Fig. 3.1 Successive absorption of wavelengths in an *n-junction* solar cell

Similarly, the second layer having bandgap of 2.50 eV can absorb only the blue photons among the all photons incident upon it. So, only the green, yellow, orange, red and infrared photons are directed towards the next subcells. This filtration process continues till the last subcell. The subcell layers are connected in series providing a higher voltage than single junction solar cells. Thus, utilizing the best photon to electricity conversion capability of each sub cell, the overall efficiency of the cell is increased.

## 3.4 Possible Structures of a Multijunction Cell

There are two methods of light distribution to the sub cells in a multijunction cell. The first method uses a beam splitting optical system to distribute sunlight to the series connected subcells and in the second method the subcells are mechanically stacked together [4, 5]. This spatial splitting can be done by tiny beam splitting filters, diffractive lenses or prisms. Since it avoids losses due to successive propagation of light as in stacked solar cells, conversion efficiency is high. However, the complexity of designing an optical splitter makes it unfavorable. The most commonly used multijunction cells are designed using the stacking technique. There are three methods to do it [7]. These methods are illustrated in figure 3.2 (b) for a double junction cell. The 2-terminal monolithic stacking is used most of the time because of the feasibility of growth process. But designing such a cell is a difficult job because of current and lattice matching requirements. The second and third of the stacking methods do not need current and lattice matching among the subcells because generated power is extracted separately from each subcell. It is very difficult to connect three-terminal devices in series, so three-terminal tandem cells do not appear to be viable. In the four-terminal case, two separate external circuit loads are used. Since the two individual cells are not coupled, the photocurrents do not have to be the same. Consequently, a much larger range of bandgap energy combinations is possible, and the changes in photocurrents with changing solar spectral distributions do not pose serious limitations. However increased number of electrodes has to be used for a four terminal structure. Losses due to shadowing of the contact metallization increase with the increase in number of electrodes. Another thing to consider is that, as power is extracted separately, individual load matching is required. Considering all these difficulties, 2-terminal stacking has become the most preferable option.

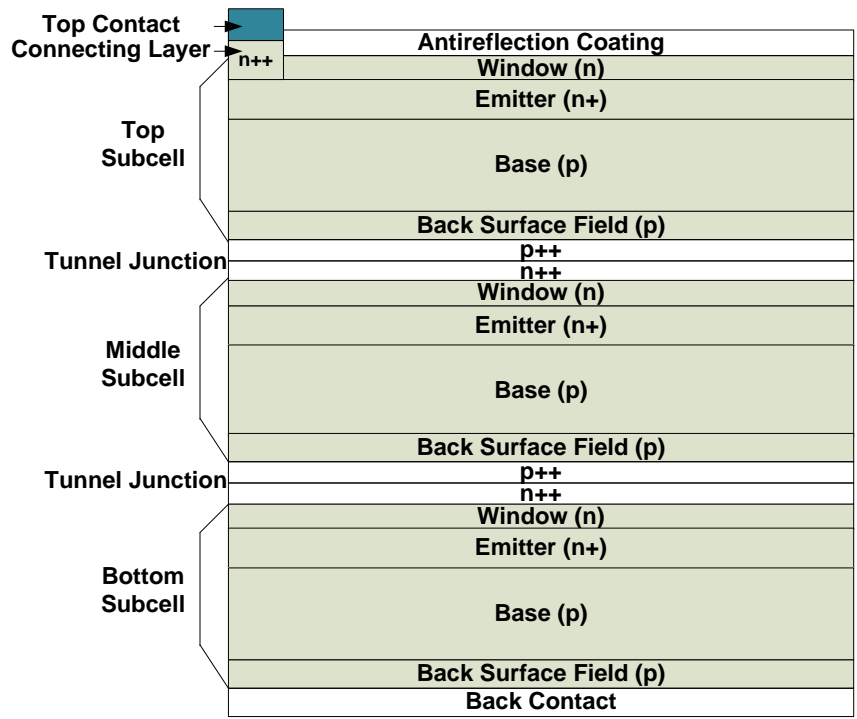

Fig. 3.3 Structure of a stacked multijunction (triple junction) cell

A triple junction cell configuration is demonstrated in figure 3.3 as an example of mechanically stacked multijunction solar cell. The top p-n junction is made of the highest bandgap material. The middle p-n junction is made of the intermediate bandgap semiconductor and the bottom p-n junction is made of the lowest bandgap semiconductor. Each of the subcells has a window layer on top of it and a back surface field layer underneath. There is a tunnel junction between every two subcells which plays a pivotal role in channeling the photons to the next layer in the stack and reducing electrical losses.

On top of the tandem of subcells, there is an antireflection coating layer. As the name suggests, anti-reflection coating reduces the reflection of light from the top surface of the solar cell. In single junction solar cells generally SiN is used to construct such a layer. However, a single layer anti-reflection coating reduces reflection well only for a portion of the total solar spectrum. A multijunction cell absorbs broader range of the solar spectrum where a SiN based coating cannot ensure zero or nearly-zero reflection. To solve this problem, a double layer antireflection coating is often used. It may be a $TiO_2+Al_2O_3$ coating or $ZnS+MgF_2$. These are higher bandgap (i.e. higher than even the top subcell) materials. Therefore they pass most of the wavelengths of sunlight. Refractive index and thickness are the most important things to consider in designing an anti-reflection coating layer.

On top of every p-n junction there is a window layer which guides light towards the junction. The materials used for window layer are crystalline in nature similar to the materials for p-n junction. Therefore it can provide low surface recombination velocity at its contact with n side (in other words, surface passivation) by fixing dangling bonds. It also acts as a selective contact which ensures unidirectional carrier flow. When an electron tries to pass through it, it allows; but a holes gets reflected back when it tries to do so. A window layer has slightly higher bandgap than the p-n junction, on top of which it is created.

A back surface field layer is similar to the window layer which is adjacent to the p side of a p-n junction. It also acts as a passivation layer for the p side. It is constructed with lower bandgap material than the p-n junction above it but higher bandgap than the p-n junction next to it. It also acts as a selective contact, but this time for hole. Any electron trying to pass through it gets reflected back.

The tunnel diodes do the important task of tunneling the carriers, otherwise they would be blocked. To illustrate the problem of blocking and its solution, let us look at figure 3.3 once again. It has been redrawn in figure 3.4. We expected creation of three diodes for a triple junction cell. However, two extra diodes have also been created with opposite polarity at the interfaces between every two junctions. These diodes block current flow in the upward direction. They are placed in between two junctions to solve this problem. With the presence of tunnel junctions, formation of the excess junctions becomes impossible. However, a tunnel junction is also in reverse bias condition which should block the current flow too. Tunnel junctions [8] are highly doped diodes (shown as p++ and n++ in figure 3.3) where depletion region is very thin which makes tunneling effect possible. In other words, current flow in reverse bias condition becomes possible. Depletion region width can be easily reduced by increasing the effective doping $N^*$ as given in equation 3.1. The peak current density in the tunnel junction, $J_p$ is also dependent upon bandgap of the constituent material and the effective doping $N^*$ by the relationship in equation 3.2 [9].

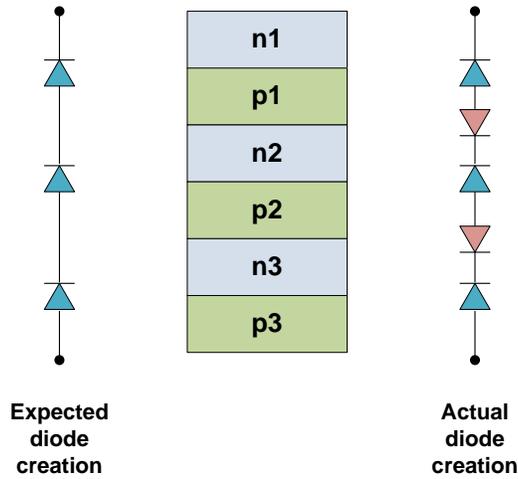

Fig. 3.4 Creation of diodes in a multijunction cell

$$W \propto \frac{1}{\sqrt{N^*}} \qquad (3.1)$$

$$J_p \propto e^{\left(\frac{-E_g^{3/2}}{\sqrt{N^*}}\right)} \qquad (3.2)$$

$$\text{where,} \qquad N^* = \frac{N_A N_D}{N_A + N_D}$$

Here, $N_A$ and $N_D$ are the doping densities in the base and emitter regions of the tunnel junction which are usually high so that the carriers with lowest biasing (by photogenerated voltage) can tunnel through it. Tunnel junctions are designed such that $J_p$ is higher than the short circuit current density of the cell, $J_{sc}$.

## 3.5 Device Performance

Multijunction solar cell tries to utilize each and every photon of sunlight and convert into electricity, this capability being limited by their design. The final throughput depends on the absorption capacity, conversion rate and losses occurred. The quantum efficiency (QE) of a subcell gives an idea about the absorption capacity of a subcell. The fraction of photons entering into a subcell which is absorbed in that subcell is called the internal quantum efficiency (IQE) of that subcell. However not all the photons incident on a subcell can enter into a subcell; some are reflected back. The fraction of photons incident on a subcell which is absorbed in that subcell is called the external quantum efficiency (EQE) of that subcell. IQE and EQE are related by the following equation:

$$IQE(\lambda) = \frac{EQE(\lambda)}{1 - Reflectivity(\lambda)} \qquad (3.3)$$

Here, we notice that absorption capability of a subcell and reflection from its surface are wavelength dependent parameters. If a very good antireflection coating (ideally which have zero reflection) is used on top of the top subcell IQE and EQE becomes same and we can express them simply by QE. The QE of a subcell can be expressed as a function of emitter, base and depletion region quantum efficiency as given in equation 3.4.

$$QE = QE_{emitter} + QE_{depletion} + \exp[-\alpha(x_e + W)]QE_{base} \qquad (3.4)$$

Here, α is called the absorption coefficient of the constituent material which is a wavelength dependent physical entity. W is the width of the depletion region and $x_e$ is the width of the emitter. Now $QE_{emitter}$, $QE_{depletion}$ and $QE_{base}$ can be expressed as follows,

$$QE_{emitter} = f_a(L_e)\left(\frac{l_e + \alpha L_e - \exp(-\alpha x_e)\left[l_e \cosh\left(x_e/L_e\right) + \sinh\left(x_e/L_e\right)\right]}{l_e \sinh\left(x_e/L_e\right) + \cosh\left(x_e/L_e\right)} - \alpha L_e \exp(-\alpha x_e)\right) \quad (3.5)$$

$$\text{where,} \qquad f_a(L_e) = \frac{\alpha L_e}{(\alpha L_e)^2 - 1}, \qquad l_e = \frac{S_e L_e}{D_e} \qquad \text{and} \qquad D_e = \frac{KT\mu_e}{e}$$

$$QE_{depletion} = \exp(-\alpha x_e)[1 - \exp(-\alpha W)] \qquad (3.6)$$

$$QE_{base} = f_b(L_b)\left(\alpha L_b - \frac{l_b \cosh\left(x_b/L_b\right) + \sinh\left(x_b/L_b\right) + (\alpha L_b - l_b)\exp(-\alpha x_b)}{l_b \sinh\left(x_b/L_b\right) + \cosh\left(x_b/L_b\right)}\right) \qquad (3.7)$$

$$\text{where,} \qquad f_b(L_b) = \frac{\alpha L_b}{(\alpha L_b)^2 - 1}, \qquad l_b = \frac{S_b L_b}{D_b} \qquad \text{and} \qquad D_b = \frac{KT\mu_b}{e}$$

Here $x_b$ is the width of the base, K is the Boltzmann constant and T is the temperature in degree Kelvin. $\mu_e$ and $\mu_b$ are the carrier mobility of electron and hole respectively. Carrier mobility is a material property which expresses how fast a carrier can move through a material. It is expressed as $cm^2/$ (V.s). $D_e$ and $D_b$ are called the diffusion constants for electron and hole respectively. The higher the diffusion constant, the faster a particle diffuses. Its unit is $cm^2$/s. The surface recombination velocity, S denotes the recombination rate in an unpassivated surface of the semiconductor. The minority carrier diffusion length, L is an important parameter for solar cell performance. It is the average length a carrier can travel between generation and recombination process. Heavy doping gives rise to longer diffusion length. Diffusion length solely depends on the material and the type of recombination occurring in it. A longer diffusion length is always expected because that prevents some recombination losses.

Surface recombination in a subcell can be prevented through the usage of passivation layers (window and back surface field). If there is no surface recombination and L is greater than the total thickness of the subcell, the set of equations from 3.5 to 3.7 results in

$$QE(\lambda) = 1 - \exp[-\alpha(\lambda)x] \qquad (3.8)$$

Here $x = x_e + W + x_b$ is the total thickness of a subcell. The short circuit current density in the subcell is given by

$$J_{SC} = e \int_0^\infty (QE(\lambda)\Phi_{inc}(\lambda)\,d\lambda) \qquad (3.9)$$

Here e is the charge of an electron and $\Phi_{inc}(\lambda)$ is the incident photon flux. This equation expresses that the photogenerated current depends not only on the design, but also on the illumination. Since the level of illumination in different regions on earth is different, performance of a cell will be different.

### 3.6 Epitaxial Technologies for Growing Multijunction Cells

The word 'epitaxy' came from two Greek roots *epi* meaning 'above' and *taxis* meaning 'an ordered manner'. Thus epitaxy is the process of depositing a crystalline overlayer on a crystalline substrate [10]. As we see there are several layers in the structure of a multijunction solar cell, so it is fabricated using epitaxy. When a layer is deposited on the substrate of the same material, it is called homoepitaxy. In homoepitaxy, doping level may be different to have different optical properties. If the overlayer is of different than that of the substrate, the deposition method is called heteroepitaxy. Although they are different materials, the overlayer and the substrate should have similar lattice structure. In case of lattice mismatch, an elastic strained structure can be produced by slowly modifying the relative properties of the material by varying its composition. This method is called grading and the substrate is called graded substrate. Liquid phase epitaxy (LPE), molecular beam epitaxy (MBE), chemical vapor deposition (also known as vapor phase epitaxy or VPE) Metal organic chemical vapor deposition (MOCVD) are the most widely used deposition method for compound semiconductors.

LPE is a simple deposition method which deposits materials in liquid phase. It uses the fact that when impurity is added to a material, its melting point decreases. Although it is a good method for growing high quality single junction solar cells, it is not efficient in growing multiple layers due to the difficulty of controlling thicknesses, doping, composition and the speed of throughput.

CVD, as its name suggests, grows crystals in gaseous form. It is done step by step. As an example, let us imagine that GaAs is to be deposited on top of a substrate. First Ga and As are transformed into gaseous state at high temperature. Then they are targeted towards the substrate at a controlled speed so that they can react with each other to form GaAs on top of the substrate. Finally the substrate is dissolved and transported out of the reaction chamber. CVD can be performed in low pressure, atmospheric pressure or high pressure conditions.

MOCVD is a special kind of CVD characterized by the chemical nature of the precursors. Metalorganic compounds with relatively high volatility are used as precursors for both, main elements and doping elements. Over time, MOCVD has been extensively used for large-scale, large-area production of multijunction solar cells for its good reproducibility and controllability.

In MBE, the material to be deposited is heated up to create an evaporated beam of molecules. This beam is then passed through a vacuum. Finally it finds a substrate as its target, on top of which it condenses and gets crystallized. MBE has lower throughput than other forms of epitaxy.

## 3.7 Cell Design Options and Issues

There are several things that have to be considered while designing a multijunction cell. The factors are described in the following subsections:

### 3.7.1 Number of Junctions

Let us reiterate the thing that, in multijunction solar cell thermalization loss occurs both in case of photon energy being higher or lower than the bandgap. As, there are countless number of photon energy levels, only an infinite junction solar cell can utilize all the energy from sunlight; which is not possible in reality. Therefore the designer has to decide the number of junctions at first. Obviously, the manufacturing cost increases with number of junctions. Figure 3.5 illustrates the relationship between efficiency and number of junctions, found using detailed balance method [11, 12]. From this figure, it is evident that efficiency (in single sun concentration) does not increase much after certain point. Therefore, for commercial applications multijunction solar cells are mostly of double or triple junctions. Quadruple junction solar cells are mostly grown in lab for research purpose only.

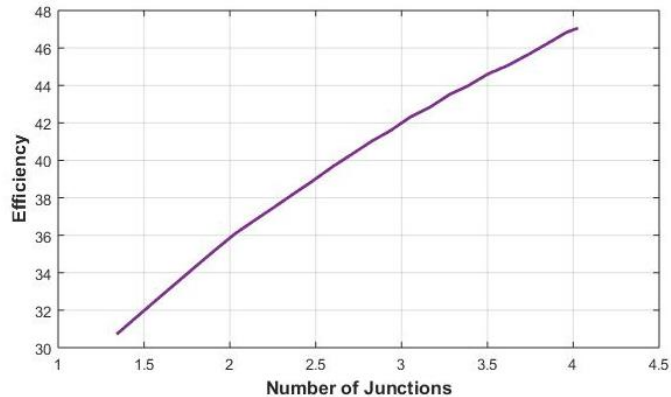

Fig. 3.5 Variation of efficiency with number of junctions using detailed balance method

### 3.7.2 Choice of Materials

Finding the best set of materials with the best composition is another challenge for the designer. This option is limited by the bandgap and lattice parameters of the materials. Both the parameters can be tuned by varying the composition of a material (bandgap engineering). It is desired that the materials have equal lattice constant values. Otherwise, threading dislocations will form due to lattice mismatch between two subcells. These dislocations will act as loss points in the solar cell. If threading dislocation exists in the bottom subcell, it may cause some other dislocations in the upper subcells at the time of growth process. Thus, finding the right material set and their compositions is very difficult. Figure 3.6 is often used as a reference for such choice.

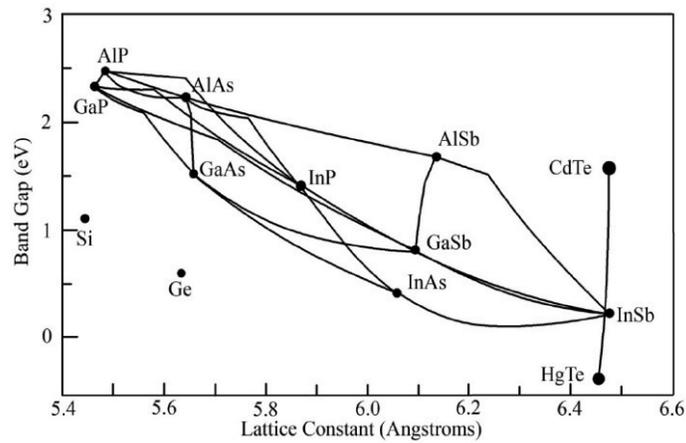

Fig. 3.6 Bandgap vs. lattice constant [13]

### 3.7.3 Growth Options

There are several methods for growing a multijunction solar cell. The most commonly used and most desired one is latticed matched growth. This type of structure mitigates the harmful effects of threading dislocations due to the lattice mismatch. Thus minimising loss, this type of cells can attain high efficiency if the right order of bandgaps is chosen. For example, a GaInP(1.75 eV)/GaInAs(1.18 eV)/Ge(0.70 eV) cell gives 40.7% efficiency at 240 sun concentration [14].

However it is not always possible to find a latticed matched material set. Fortunately, technological advancements allow us to grow slightly lattice mismatched subcells on top of each other. This type of growth process is called upright metamorphic (or simply metamorphic) growth. In this process, at first the lowest bandgap material is deposited on top of the substrate. Then other materials having higher bandgap are deposited one by one.

Stacking faults, point defect etc can be the loss centers in a metamorphic design. Losses occur due to the recombination in these defect centers. Even the simplest defect in the lower subcells in a metamorphic design can turn into a bigger one because it propagates through the upper subcells. To avoid these problems another approach named inverted metamorphic growth has been popular recently. Unlike the metamorphic design, in this method the highest bandgap material is deposited on top of the substrate at first. Then the second highest bandgap material is deposited on top of it. This process continues till the lowest bandgap material is deposited. After the deposition is done, the substrate is removed and the entire cell is flipped so that light can enter from the highest bandgap side.

There is another method called wafer bonding. In this method two separately prepared pieces of a cell are wafer bonded with each other. Thus it avoids the creation of threading dislocations. However, the wafer bonding creates a lot of unpassivated areas with higher surface recombination velocities. This causes recombination losses in the surface.

### 3.7.4 Difference in Solar Spectrum

Complexity of design grows more due to the difference in solar intensity in different places on earth and also in space. The flux of sunlight in Arizona is surely different than that in Tennessee. To understand the problem, let us imagine that a multijunction solar cell is designed for AM 1.5 solar spectrum. Current through all the cells is matched and thickness and doping of the individual subcells are optimized. If we use this solar cell in space, we will not get the same efficiency as that obtained on earth. In the changed illumination, current density in the subcells will no longer be the same; some subcells will have lower current and some will have higher than the previous time. As all the subcells are in series, the lowest amount of current will be the resultant current. The excess current than this common current value will generate heat. It will degrade the cell performance further. Therefore, a cell has to be redesigned for the changed situation.

## 3.8 Key to Realizing High Efficiency Multijunction Cell

In the complex task of designing a multijunction solar cell, there are several ways to enhance the efficiency of a cell. These are discussed in the following subsections:

### 3.8.1 Ensuring Enough Minority Carrier Lifetime & Diffusion Length

Minority carrier lifetime and diffusion length depend on the recombination tendency of the constituent material. The defects and impurities in the material often act as recombination center. If minority carrier lifetime or diffusion length is short, the photogenerated minority carriers cannot be collected because they recombine very fast. The minority carrier lifetime can be expressed as [15],

$$\tau = \frac{1}{BN} \qquad (3.10)$$

Here N is the carrier concentration and B is the radiative recombination coefficient. N can be optimized by considering the built-in potential and series resistance of the subcells. Choosing is proper material for the top subcell is also important in this regard. GaInP is better than AlGaAs in the sense that in AlGaAs has residual oxygen in it which acts as recombination center.

### 3.8.2 Wide Bandgap Window and Back Surface Field

The window and back surface field (BSF) layers are used as passivation layers. However, short circuit current density decreases with increase in their thickness [15]. So, they should be as thinner as possible. The constituent materials should also be of wide bandgap so that they cannot absorb photons. A heteroface or double-hetero structure window layer is effective.

### 3.8.3 Low Loss Tunnel Junction

The tunnel junctions which interconnect two adjacent subcells should cause low optical and electrical losses. A degenerately doped tunnel junction is attractive because creating it only involves one extra step in the growth process. These tunnel junctions should be wide band gap but physically thin so that they cannot reduce the current density in the cell. However tunneling current decreases exponentially with increase in bandgap energy [15]. To counteract this effect, the tunnel junctions should be highly doped.

# CHAPTER 4

# NOVEL QUADRUPLE JUNCTION SOLAR CELL DESIGN

## 4.1 Objectives of the Design

The main objective of this thesis was to design a multijunction solar cell which can absorb photons in most of the solar spectrum and thus provides high light to electricity conversion efficiency. Considering design complexity, fabrication cost and throughput from a solar cell, designing a four junction solar cell (no more or less than four junctions) seemed appropriate. The present quadruple junction solar cell record efficiency is 38.8% under single sun AM 1.5 solar irradiation and 46.0% with concentrated light. The main objective was to surpass this efficiency limit in the novel design. As the same current density should be generated in all of the four junctions, compositions of the materials should be carefully chosen to ensure the appropriate combination of bandgaps. Materials with low surface recombination velocity, low lattice mismatches and longer carrier lifetimes are of great interest for the design to ensure reduced recombination and thermalization losses. The novel quadruple junction solar cell comprising GaInP/GaAs/InGaAs/InGaSb subcell layers provides 47.20% power conversion efficiency.

## 4.2 Introduction

The inability of single junction solar cells in absorbing the whole solar spectrum efficiently has led the researchers to multijunction approach. A multijunction solar cell consists of several subcell layers (or junctions), each of which is channeled to absorb and convert a certain portion of the sunlight into electricity. Each subcell layer works as a filter, capturing photons of certain energy and channel the lower energy photons to the next layers in the tandem. The subcell layers are connected in series providing a higher voltage than single junction solar cells. Thus, utilizing the best photon to electricity conversion capability of each subcell, the overall efficiency of the cell is increased [1].

There are two methods of light distribution to the subcells in a multijunction cell. The first method uses a beam splitting filter to distribute sunlight to the series connected subcells and in the second method the subcells are mechanically stacked together [1]. The portion of the solar spectrum a subcell will absorb depends on the bandgap of the material used. Higher bandgap materials absorb higher energy photons and give relatively higher amount of voltage. Since number of higher energy photons is limited, number of excitons (electron-hole pair) generated and current is limited. On the contrary, materials with lower bandgap absorb lower to higher energy photons and give lower voltage but higher current. Therefore, choosing an appropriate set of high to low bandgap materials is important in multijunction solar cell design. This job can be challenging because the adjacent subcells should also be lattice matched to minimize threading dislocations [2].The presence of dislocation reduces the open circuit voltage ($V_{oc}$) and hence the overall power conversion efficiency of the solar cell [3]. Fortunately, there are some technologies that allow lattice mismatch up to certain limits. Metamorphic design uses buffers to limit formation of dislocations [4]. Inverted metamorphic technology is a modified version of metamorphic technique where some cells are at first grown on a temporary parent substrate. The cells are then placed on the final substrate upside down and the temporary parent substrate is removed [5]. Direct wafer bonding is another way which forms atomic bonds between two lattice mismatched materials at their interface and thus eliminates the dislocations [6]. Some authors utilized this method successfully to address relatively higher mismatch value like 3.7% and 4.1% [7, 8].

After choosing the appropriate materials, current matching becomes the most important task in the design procedure. Since the subcells are connected in series, the lowest current density determines the overall current density of the cell. If current values are not matched, the excess current in the subcells other than the subcell with lowest current density gets lost as heat. The impact is twofold: firstly, some energy is lost; secondly, the heat generated deteriorates the cell performance further.

Solar cell is an excellent renewable power source. However, higher conversion efficiency and cost-effectiveness have been the major issues for large scale commercial applications of these cells. Theoretically a multijunction solar cell can give us 86.4% power conversion efficiency with infinite number of junctions [9]. Of course, manufacturing cost increases if higher numbers of junctions are used. When cost is an important factor in determining the market share of solar modules in the current power sector, we want to design a solar cell which has lesser number of junctions but gives relatively higher efficiency. The calculations using detailed balance method shows that, the highest efficiency achievable from a quadruple junction solar cell is 47.5% for single sun condition and 53% for maximum concentration of sunlight [3, 10]. This theoretical approach assumes ideal cases i.e. no reflection loss, zero series resistance of subcells and tunnel junctions, 300K temperature and no re-absorption of emitted photons [1]. However, the highest practical efficiency achieved till now is only 46.0% which assembled four subcells with concentrators [12]. For 1-sun condition the efficiency is noticeably lower; 38.8% using five subcells [12].

Solar energy ranges from ultraviolet to infrared region. Previously InGaP/GaAs/InGaAs [13] based triple junction solar cell was proposed which cannot capture much in the infrared region. To utilize infrared portions too, Ge was used as a bottom subcell layer [3, 11, 14]. Bhattacharya et al. proposed another material, InGaSb which is good at capturing infrared photons [15]. It was used in GaP/InGaAs/InGaSb based triple junction solar cell later on [16, 17]. In this thesis, an $In_{0.51}Ga_{0.49}P/GaAs/In_{0.24}Ga_{0.76}As/In_{0.19}Ga_{0.81}Sb$ based quadruple junction solar cell is being proposed for the first time. The electronic bandgap of these materials are 1.9 eV, 1.42eV, 1.08 eV and 0.55 eV respectively which help proper distribution of light to all the junctions. The first two junctions are lattice matched. Lattice mismatch between GaAs and InGaAs is 2.78% when it is 5.59% between $In_{0.24}Ga_{0.76}As$ and $In_{0.19}Ga_{0.81}Sb$. Appropriate fabrication technique like metamorphic, inverted metamorphic or wafer bonding needs to be used to make the structure defect free. The simulation result shows that current density is same in all the junctions. This reduces the possibility of energy loss and performance deterioration. The theoretical power conversion efficiency of the cell is 47.2082%. This value is higher than the present record efficiency quadruple junction solar cell with concentrators (46.0%) [12].

### 4.3 Proposed Quadruple Junction Solar Cell

Material selection with proper bandgap is an important factor in designing high efficiency multijunction solar cell. III-V compound semiconductors are generally chosen because of their bandgap tunability through elemental composition. These compound semiconductor alloys have band gaps ranging from 0.3 to 2.3 eV which cover most of the solar spectrum [1]. The proposed novel quadruple junction cell is also designed from III-V compounds, comprising InGaP, GaAs,

InGaAs and InGaSb subcell layers respectively. Key features of the design are discussed below:

### 4.3.1 Structure

The quadruple junction solar cell consists of four sub cells connected in series, as shown in Fig. 1. Each subcell has three parts: n type emitter, p type base and a back surface field (BSF) layer. Base is made thicker than emitter because of the work function of p type base being higher than n type emitter layer.

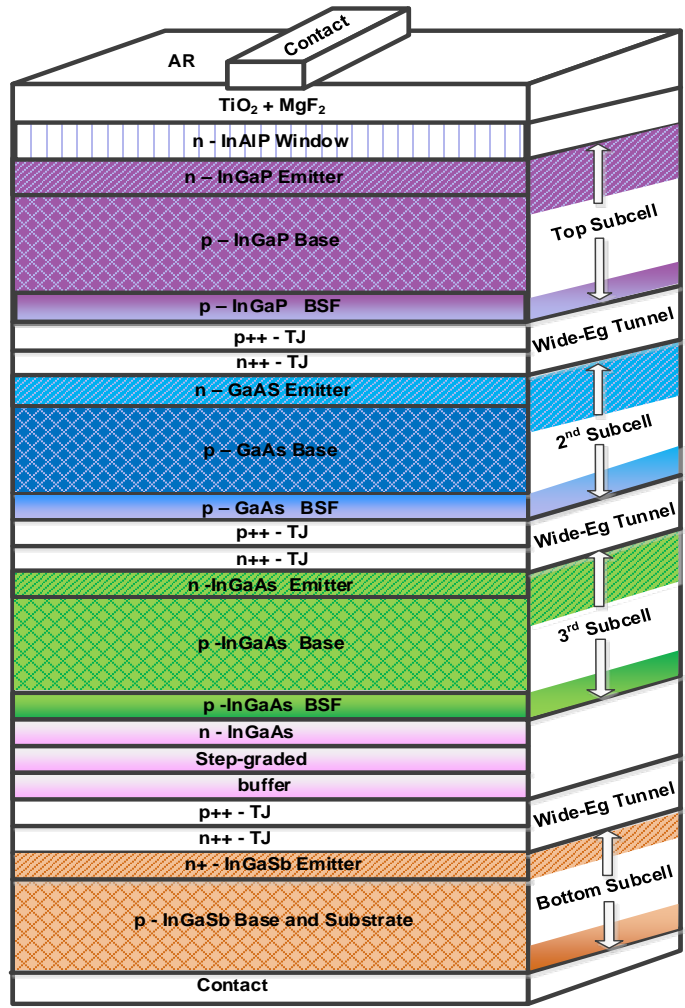

Fig 4.1 Structure of the novel quadruple junction solar cell

The electron-hole pairs (excitons) are generated in the p-n junction formed in the interface between emitter and base which contributes to the photocurrent. The back surface field is made of the same material. It fixes dangling bonds and thus reduces surface recombination. Two adjacent subcells are connected together by tunnel diodes. Higher level of doping is used to design these tunnel diodes which help them not absorb light and exhibit tunneling effect. Antireflection (AR) coating is a special type of layer used to reduce reflection of light fallen on the solar cell. With double layer $TiO_2$+$MgF_2$ antireflection coating, reflection loss can be reduced to 1%. The window layer acts as a means of light passage to the p-n junction. It protects the cell from outside hazards too. Step graded buffers are used to eliminate the threading dislocations formed between $In_{0.24}Ga_{0.76}As$ and $In_{0.19}Ga_{0.81}Sb$ lattice mismatched sub cells. The front and back contacts are used to collect photocurrent from the solar cell.

### 4.3.2 Material Properties

The material properties considered for the design are summarized in Table 4.1. Most of the properties are temperature dependent. All through the design process we considered 300K temperature. The top subcell is made of a high bandgap material, $In_{0.51}Ga_{0.49}P$ with bandgap of 1.9 eV [18]. This enables it to absorb photons in the ultraviolet region efficiently. GaAs has bandgap of 1.42 eV which empowers it to absorb most of the sunlight in visible range. The bandgap of $In_{1-x}Ga_xAs$ is $(0.36+0.63x+0.43x^2)$ eV [19]. With x=0.76, it becomes 1.08 eV. The bottom sub cell is made of low bandgap material, $In_{0.19}Ga_{0.81}Sb$ whose bandgap may be expressed as, $Eg= (0.7137-0.9445x+0.3974x^2)$ eV [20], where x is the indium composition. With x=0.19, bandgap becomes 0.55 eV. Due to this lower bandgap value it can absorb in infrared region. The doping level of emitter is higher than base. Window layer is normally made of higher bandgap and highly doped n type material. Due to the high doping used and very little thickness, it does not absorb any photon and passes light to the subcells next in the tandem. The doping level of tunnel junction is even higher. The lattice constant of a material also depends on the composition. GaAs has a lattice constant of 5.65325 Å [21].The general expressions of lattice constants for $In_{1-x}Ga_xP$, $In_{1-x}Ga_xAs$ and $In_{1-x}Ga_xSb$ are (5.8687-0.4182$x$) Å, (6.0583-0.405$x$) Å and (6.479-0.383x) Å respectively [22-24]. The values become 5.653 Å, 5.8153 Å and 6.16 Å for $In_{0.51}Ga_{0.49}P$, $In_{0.24}Ga_{0.76}As$ and $In_{0.19}Ga_{0.81}Sb$ respectively. Since all these four materials have the same Zinc Blende crystal structure [21-24], defects occurred from the lattice mismatch can be easily eliminated by adopting appropriate technology i.e. metamorphic, inverted metamorphic, wafer bonding etc. Step graded buffers used in this structure solves the dislocation problem further. Minority carrier lifetime is another important parameter. If it is very low then some of the photocurrents are lost before they can be collected. It is in the order of $10^{-3}$ s for $In_{0.51}Ga_{0.49}P$ and $10^{-8}$ s for GaAs [25]. For $In_{0.24}Ga_{0.76}As$, carrier lifetime depends on doping level through the relation, $\tau = (2.11*10^4 + 1.43*10^{-10}*N + 8.1*10^{-29}*N^2)^{-1}$ s [26], where N is the doping density and $\tau$ is the carrier lifetime.

Front and back contacts are made of metals having very low resistances so that they can collect the generated photocurrent without any loss. The doping concentration for emitter of each subcell was designed to be in the order of $10^{18}/cm^3$. The highest value of doping concentration for base is in the order of $10^{17}/cm^3$. Surface recombination velocities of the materials used are in the order of $10^5$ cm/s [27-29]. Therefore recombination losses were considered in the design.

Table 4.1: Material Properties Assumed for the Design

| Material Properties | | Top subcell ($In_{0.51}Ga_{0.49}P$) | Subcell-2 (GaAs) | Subcell-3 ($In_{0.24}Ga_{0.76}As$) | Bottom Subcell ($In_{0.19}Ga_{0.81}Sb$) |
|---|---|---|---|---|---|
| Bandgap (eV) | | 1.9 | 1.42 | 1.08 | 0.55 |
| Lattice Constant (Å) | | 5.653 | 5.65325 | 5.8153 | 6.16 |
| Intrinsic Carrier Concentration (/cm$^3$) | | $1*10^3$ | $1.79*10^5$ [30] | $1.31*10^9$ | $2.5*10^{13}$ |
| Surface Recombination velocity (cm/s) | | $4*10^5$ | $5*10^5$ | $1*10^4$ | $0.5*10^5$ |
| Dielectric Constant | | 11.8 | 12.9 | 13.3058 | 16 |
| Diffusion Coefficients | Electron | 26.8 | 200 [31] | 220 [32] | 297.7030 [33] |
| | Hole | 3.8 | 0.5 [31] | 0.09 [32] | 0.5170 [33] |
| Minority Carrier Lifetime (s) | Electron | $0.1*10^{-3}$ | $10^{-8}$ [34] | $1.3562*10^{-7}$ | $9*10^{-9}$ |
| | Hole | $0.1*10^{-3}$ | $10^{-8}$ [34] | $1.4149 \times 10^{-10}$ | $9*10^{-9}$ |
| Doping (/cm$^3$) | Emitter | $8.5*10^{18}$ | $3.5*10^{18}$ | $8.5*10^{18}$ | $8.5*10^{18}$ |
| | Base | $3.5*10^{17}$ | $1.1*10^{15}$ | $5*10^{16}$ | $3.5*10^{17}$ |

The materials have been carefully chosen so that lattice mismatch between two adjacent subcells is low. This made the design more challenging; choice of materials and their composition became narrow due to bandgap-lattice constant tradeoff.

## 4.4  Design Approach

To design the quadruple junction solar cell we made some assumptions that are generally done for simplification in solar cell modeling. These assumptions are [35]: transparent tunnel junction interconnects with no resistance, no reflection loss and no series resistance loss in the junctions and p-n junctions formed are ideal (diode ideality factor, n is equal to 1). According to these assumptions, if a photon is absorbed by a subcell, one exciton (electron-hole pair) is generated. The fraction of the total number of photons absorbed in a subcell is determined by the thickness ($x_i$) of that subcell and the absorption coefficient ($\alpha$) of the constituent material. For the design, the absorption data of $In_{.51}Ga_{0.49}P$, GaAs, $In_{0.24}Ga_{0.76}As$ and $In_{0.19}Ga_{0.81}Sb$ were collected from [18], [36], [37] and [20] respectively.

### 4.4.1 Quantum Efficiency

Global AM 1.5 solar spectrum was considered for photon flux incident on the solar cell. The top subcell absorbs a portion of this incident photon flux. The rest is transmitted to the next subcells. Thus, the photons incident on a subcell depends on the properties of the other subcells stacked above it in the tandem. If $\emptyset_s$ is the photon flux falling on the top subcell, the amount incident on any m$^{th}$ subcell lying below, $\emptyset_m(\lambda)$ can be expressed as equation 4.1. The percentage of absorbed photons converted into electron-hole pair in a subcell is called internal quantum efficiency (QE) of that subcell. It depends on absorption coefficient $\alpha(\lambda)$, base thickness $x_b$, emitter thickness $x_e$, depletion width $W$, base diffusion length $L_b$, emitter diffusion length $L_e$, surface recombination velocity in base $S_b$, surface recombination velocity in emitter $S_e$, base diffusion constant $D_b$ and emitter diffusion constant $D_e$, as given in equation 4.2 through 4.8.

$$\emptyset_m(\lambda) = \emptyset_s(\lambda) exp[-\sum_{i=1}^{m-1} \alpha_i(\lambda) x_i] \tag{4.1}$$

$$QE = QE_{emitter} + QE_{depl} + QE_{base} \times \exp(-\alpha(x_e + W)) \tag{4.2}$$

$$QE_{depl} = \exp(-\alpha x_e)[1 - \exp(-\alpha W)] \tag{4.3}$$

$$QE_{emitter} = f_\alpha(L_e) \left( \frac{l_e + \alpha L_e - \exp(-\alpha x_e) \times \left[l_e \cosh\left(\frac{x_e}{L_e}\right) + \sinh\left(\frac{x_e}{L_e}\right)\right]}{l_e \sinh\left(\frac{x_e}{L_e}\right) + \cosh\left(\frac{x_e}{L_e}\right)} - \alpha L_e \exp(-\alpha x_e) \right) \tag{4.4}$$

$$QE_{base} = f_\alpha(L_b) \left( \alpha L_b - \frac{l_b \cosh\left(\frac{x_b}{L_b}\right) + \sinh\left(\frac{x_b}{L_b}\right) + (\alpha L_b - l_b) \exp(-\alpha x_b)}{l_b \sinh\left(\frac{x_b}{L_b}\right) + \cosh\left(\frac{x_b}{L_b}\right)} \right) \tag{4.5}$$

$$l_b = \frac{S_b L_b}{D_b} \quad (4.6) \qquad f_\alpha(L) = \frac{\alpha L}{(\alpha L)^2 - 1} \quad (4.7) \qquad l_e = \frac{S_e L_e}{D_e} \quad (4.8)$$

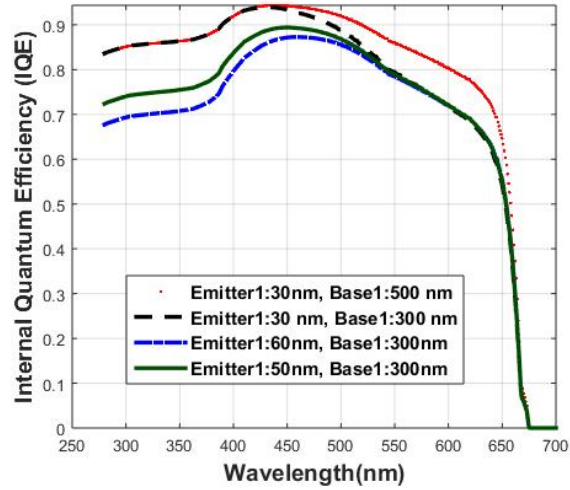

Fig. 4.2 (a) Change of quantum efficiency with change in thickness (Top subcell-InGaP)

From equation 4.2, it is evident that the quantum efficiency of emitter, base and depletion region, all contribute to the overall quantum efficiency of the cell. Among the deciding factors of quantum efficiency, absorption coefficient, surface recombination velocity, diffusion length etc. are material properties which cannot be tuned once a particular material is chosen. However thickness can be easily varied in design process to obtain the highest possible quantum efficiency. The two quantities $x_e/L_e$ and $x_b/L_b$ are significant in the expression for emitter and base quantum efficiency. This gives an idea that the capability of tuning quantum efficiency by changing thickness is limited by the diffusion length of the material used. Since we assumed no reflection loss due to the usage of double layer antireflection coating, internal quantum efficiency equals the external quantum efficiency. Changes in the quantum efficiency values with the change in thickness was investigated and illustrated in Fig. 4.2(a) through 4.2(d).

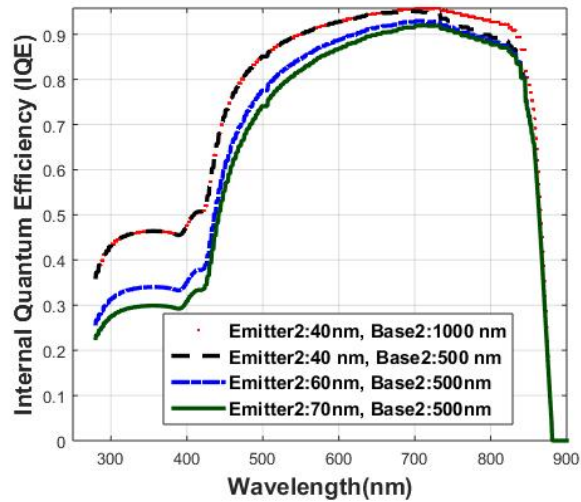

Fig. 4.2 (b) Change of quantum efficiency with change in thickness (Second subcell-GaAs)

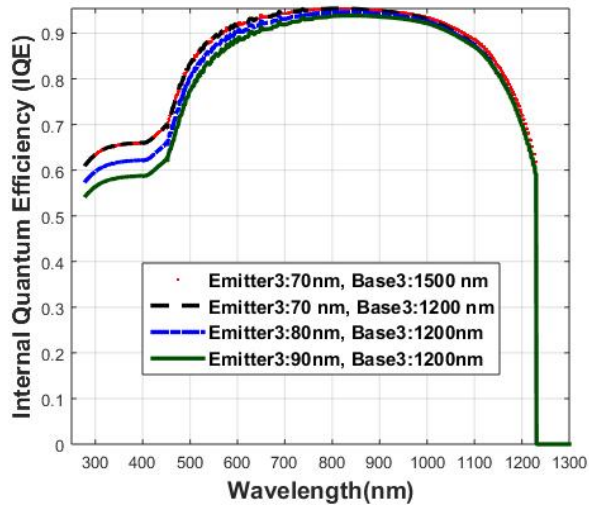

Fig. 4.2 (c) Change of quantum efficiency with change in thickness (Third subcell-InGaAs)

We noticed that quantum efficiency increases with increase in base thickness up to a certain limit. After that, an increase in base thickness has no or little impact on quantum efficiency. Increase in emitter thickness on the contrary decreases quantum efficiency in most cases. The reason behind this is, work function for hole is greater than the electron.

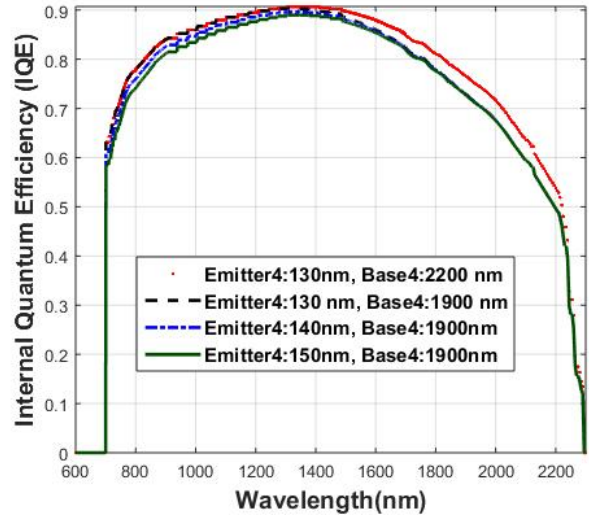

Fig. 4.2 (d) Change of quantum efficiency with change in thickness (Fourth subcell-InGaSb)

If emitter (n type) thickness increases, it absorbs some extra energy that would otherwise be absorbed in base (p type). Thus hole generation being impeded, quantum efficiency decreases.

### 4.4.2 Current Density

The short circuit photocurrent density, $J_{sc}$ obtained in a subcell depends on the quantum efficiency and the photon flux $\emptyset_{inc}$ incident on that subcell as follows,

$$J_{SC} = e\int_0^\infty (QE(\lambda)\Phi_{inc}(\lambda)\,d\lambda) \qquad (4.9)$$

Here e is the charge of an electron (1.6×10$^{-19}$ C). The incident photon flux $\emptyset_{inc}$ dependson the order of the subcell and geometry of the sub cells above, as given in equation (1). In a solar cell, photocurrent is generated due to the minority electrons in the base and the minority holes in the emitter. Little amount of reverse current is also generated due to the majority carriers, which is a loss for solar cell. This current density is called dark current density ($J_0$).The photogenerated open circuit voltage can be written as,

$$V_{OC} \approx (kT/e)\ln(J_{SC}/J_o) \qquad (4.10)$$

Where $K$ is the Boltzmann's constant and $T$ is the temperature in degree Kelvin. Using the diode characteristic equation, we determine the effective photocurrent of a subcell as,

$$J = J_o \left[\exp\left(eV/nK_BT\right) - 1\right] - J_{SC} \qquad (4.11)$$

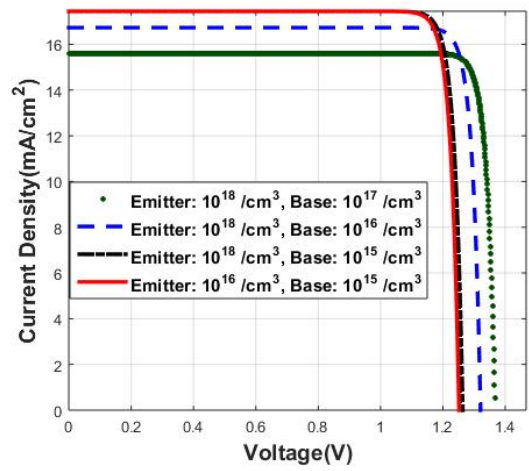

Fig. 4.3 (a) Change of J-V curve with change in doping (Top subcell-InGaP)

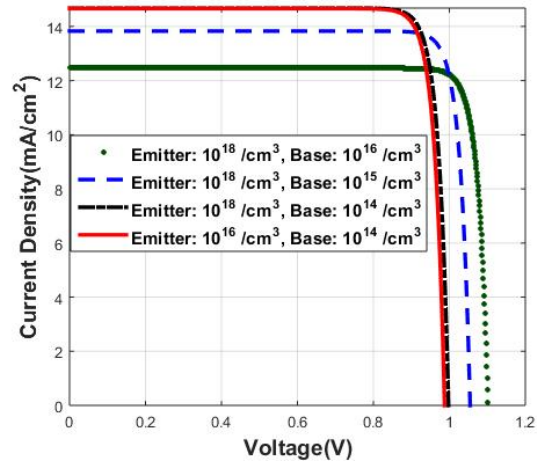

Fig. 4.3 (b) Change of J-V curve with change in doping (Second subcell-GaAs)

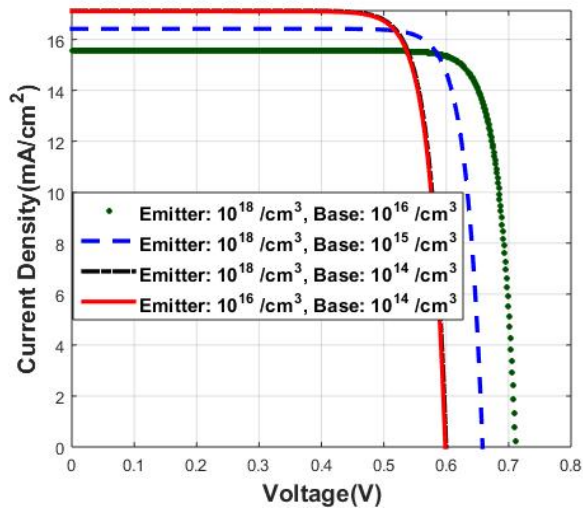

Fig. 4.3 (c) Change of J-V curve with change in doping (Third subcell-InGaAs)

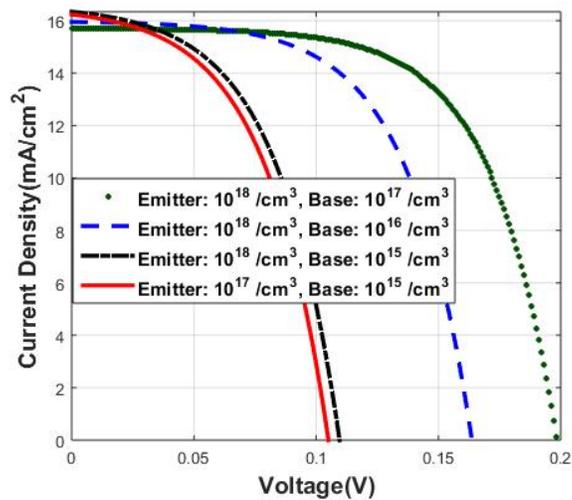

Fig. 4.3 (d) Change of J-V curve with change in doping (Fourth subcell-InGaAs)

The impact of change in doping on J-V curves were investigated and illustrated in figure 4.3. It can be noticed that increase in emitter doping increases the voltage a little bit. However the same increase for base reduces the current little bit and increases the voltage considerably. We know, p side is thicker and n side is relatively thinner in a solar cell. Thus doping in p type base evokes greater consequences. Increase in doping increases the distance between quasi Fermi levels, $E_{FP}$ and $E_{FN}$. As we know, photo-voltage is proportional to ($E_{FN}$ - $E_{FP}$), voltage increases with increase in doping. Similarly, since ($E_{FN}$ - $E_{FP}$) increases, the incident photon now requires slightly higher energy to produce an electron-hole pair. Since lesser number of photons is able to attain that energy, amount of current gets reduced.

### 4.4.3 Voltage

In a multijunction solar cell all the subcells are connected in series. Therefore, current matching is very important. If current density in all the subcells are not matched, the excess current in a subcell, being unable to flow, will be lost as heat. This thermalization will also give rise to deteriorated cell performance. In a current matched cell, the current density of the overall cell is the current density of any particular subcell, $J$. Also, the total open circuit voltage is the sum of the voltages in the subcells.

$$V_{total} = \sum_{i=1}^{m} V_i \quad (12)$$

### 4.4.4 Fill Factor

Fill factor for a solar cell can be empirically expressed as [38],

$$FF = \frac{V_{OCnormalised} - \ln(V_{OCnormalised} + 0.72)}{V_{OCnormalised} + 1} \quad \text{Where,} \quad V_{OCnormalised} = \frac{e}{nkT} V_{OC} \quad (13)$$

### 4.4.5 Efficiency

Finally, the power conversion efficiency of a solar cell is,

$$\eta = \frac{J_{sc} \times V_{oc} \times FF}{P_{in}} \times 100\% \tag{14}$$

Here $P_{in}$ is the input power (sunlight) to the solar cell. In standard test case it is 1000 W/m² for global AM 1.5 solar spectrum. For extraterrestrial illumination (AM 0) input power is 1353 W/m². The numerator expresses the power generated (electricity) from the solar cell per square meter.

### 4.5 Optimization of the Design

Doping and thickness value of the subcells were tuned to achieve the highest efficiency possible. The optimization trial is given in table 2. In the first design all the base (p type) doping were kept in the order of $10^{17}$/cm³ and emitter in the order of $10^{15}$ and $10^{16}$/cm³. Emitter thickness values were set to 45, 65, 95 and 150 nm for first, second, third and fourth subcell respectively. Base thicknesses were set to 220, 700, 1540 and 2820 nm respectively. With this arrangement, 44.1055% power conversion efficiency was found. Doping level was increased in the second design. Thickness values were also changed accordingly to match the short circuit current density at 14.7 mA/cm². This led to the increase of efficiency value to 45.3177%. Thickness values were changed in design 3 keeping the doping level unchanged, except in base of subcell 2. With this trial open circuit voltage decreased little bit. However, the considerable increase in 2mA/cm² current contributed to the increased efficiency of 46.0251%. Finally, both doping and thickness values were tuned to different values.

Table 4.2: Design Optimization for the Highest Efficiency

| Parameters | | Design 1 | Design 2 | Design 3 | Optimized Design |
|---|---|---|---|---|---|
| Doping Density (/cm$^3$) | Emitter 1 | 6.5×10$^{17}$ | 8.5×10$^{18}$ | 8.5×10$^{18}$ | 8.5×10$^{18}$ |
| | Base 1 | 3.5×10$^{16}$ | 7.5×10$^{16}$ | 7.5×10$^{16}$ | 3.5×10$^{17}$ |
| | Emitter 2 | 3.5×10$^{17}$ | 3.5×10$^{18}$ | 3.5×10$^{18}$ | 3.5×10$^{18}$ |
| | Base 2 | 0.1×10$^{15}$ | 0.3×10$^{15}$ | 0.4×10$^{15}$ | 1.1×10$^{15}$ |
| | Emitter 3 | 8.5×10$^{17}$ | 8.5×10$^{18}$ | 8.5×10$^{18}$ | 8.5×10$^{18}$ |
| | Base 3 | 0.2×10$^{15}$ | 0.7×10$^{15}$ | 0.7×10$^{15}$ | 1.5×10$^{16}$ |
| | Emitter 4 | 9.0×10$^{17}$ | 9.0×10$^{18}$ | 9.0×10$^{18}$ | 8.5×10$^{18}$ |
| | Base 4 | 8.5×10$^{15}$ | 8.5×10$^{16}$ | 8.5×10$^{16}$ | 3.5×10$^{17}$ |
| Thickness (nm) | Emitter 1 | 45 | 45 | 30 | 30 |
| | Base 1 | 220 | 300 | 270 | 400 |
| | Emitter 2 | 65 | 65 | 55 | 40 |
| | Base 2 | 700 | 900 | 700 | 1310 |
| | Emitter 3 | 95 | 90 | 70 | 70 |
| | Base 3 | 1540 | 1540 | 1460 | 1870 |
| | Emitter 4 | 150 | 150 | 120 | 140 |
| | Base 4 | 2820 | 2220 | 2200 | 2200 |
| Voltage (V) | Subcell 1 | 1.3313 | 1.3596 | 1.3579 | 1.4012 |
| | Subcell 2 | 0.9892 | 1.0236 | 1.0251 | 1.0663 |
| | Subcell 3 | 0.6152 | 0.6415 | 0.6413 | 0.6635 |
| | Subcell 4 | 0.1627 | 0.2173 | 0.2130 | 0.2422 |
| Matched Current, J$_{sc}$ (mA/cm$^2$) | | 15.0 | 14.7 | 14.9 | 14.7 |
| Open Circuit Voltage, V$_{oc}$ (V) | | 3.0984 | 3.2420 | 3.2419 | 3.3731 |
| Fill Factor (FF) | | 0.9521 | 0.9538 | 0.9538 | 0.9553 |
| Efficiency (%) | | 44.1055 | 45.3177 | 46.0251 | 47.2082 |

This step resulted in 47.2082% efficiency with short circuit current density of 14.7 mA/cm$^2$, open circuit voltage of 3.3731 V and fill factor of 0.9553.

No further improvement in efficiency was possible by changing doping and thickness values of the emitter and base layers of the subcells. Fig 4.4 (a) through 4.4 (h) illustrate the change in efficiency of the novel quadruple junction solar cell with corresponding change in doping concentrations. The changes in efficiency with corresponding changes in thickness values are also depicted in Fig 4.5 (b) through 4.5 (h). A very little improvement in efficiency is noticed for doping or thickness values other than the optimized design sometimes; however, current matching was not obtained for those designs and hence should not be adopted due to significant losses that may occur through thermalization.

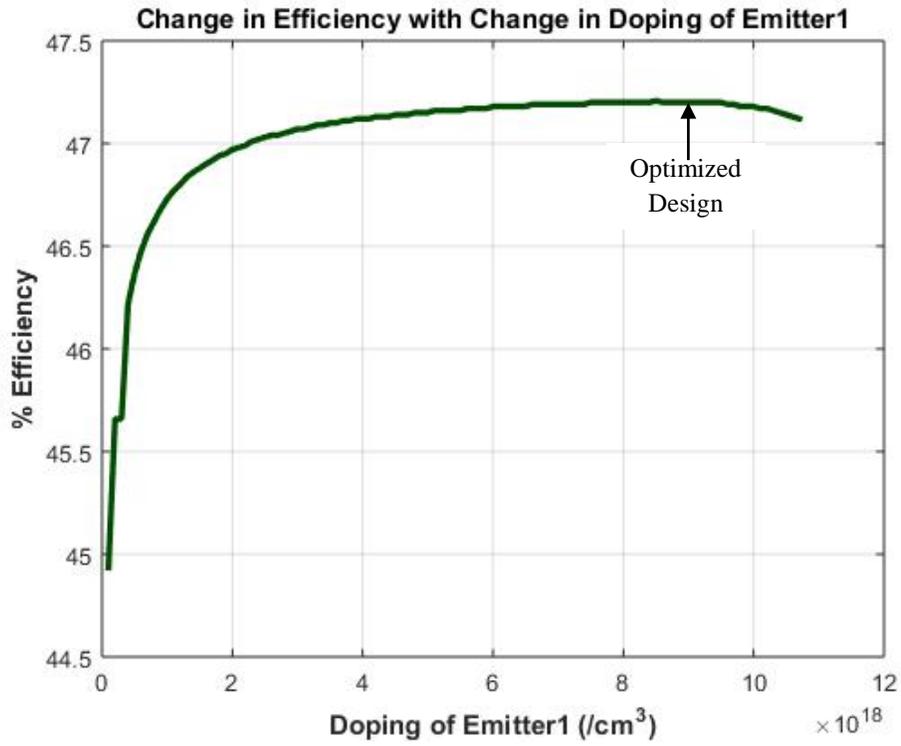

4.4 (a) Efficiency change with change in emitter1 doping

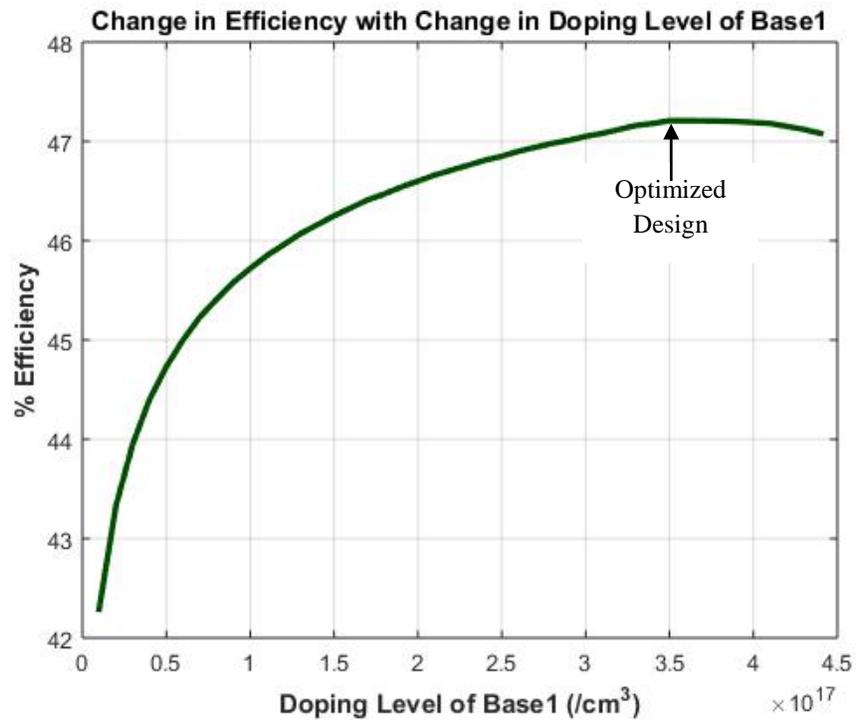

4.4 (b) Efficiency change with change in base1 doping

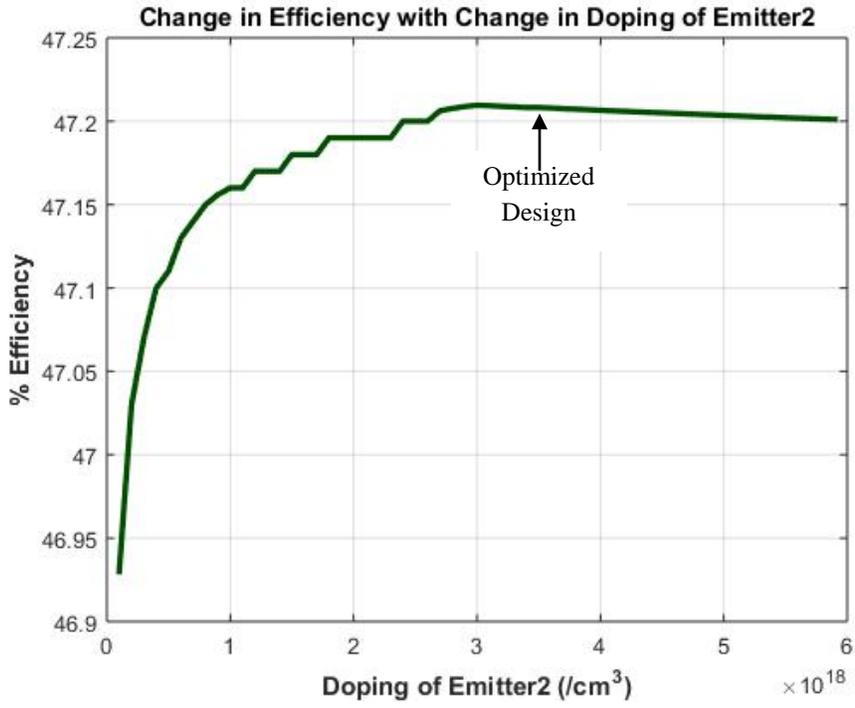

4.4 (c) Efficiency change with change in emitter2 doping

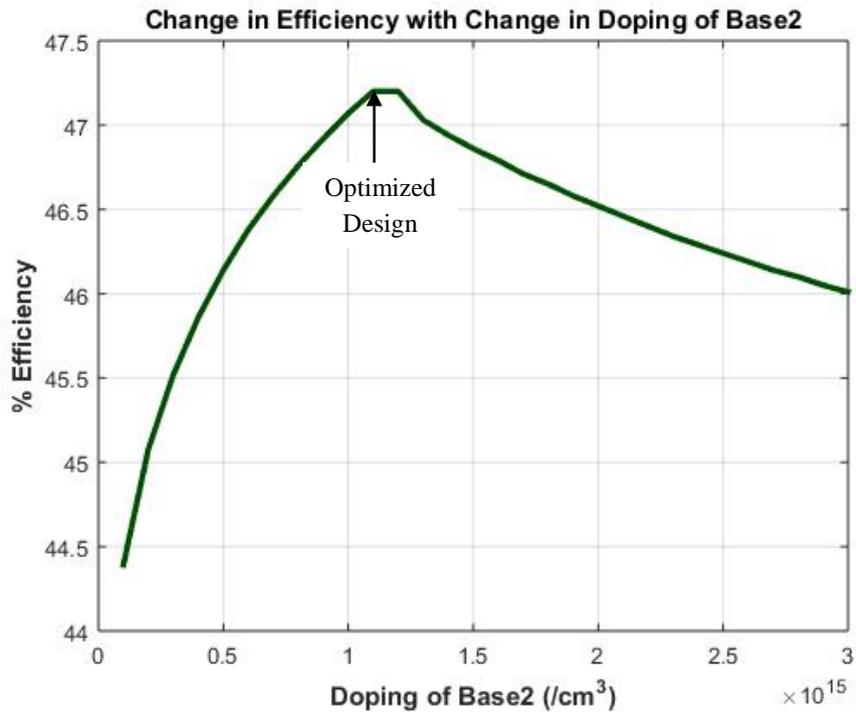

4.4 (d) Efficiency change with change in base2 doping

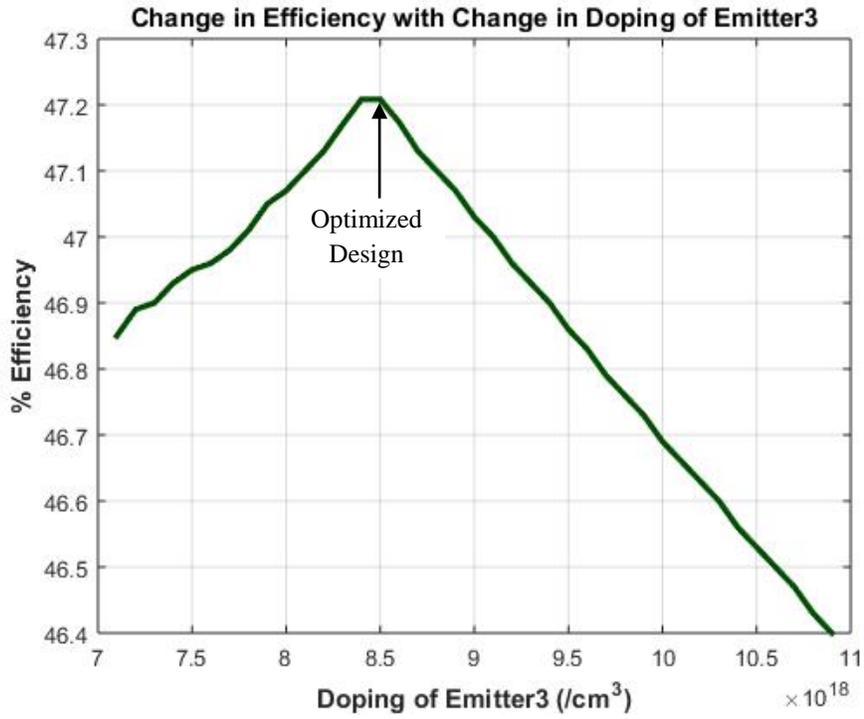

4.4 (e) Efficiency change with change in emitter3 doping

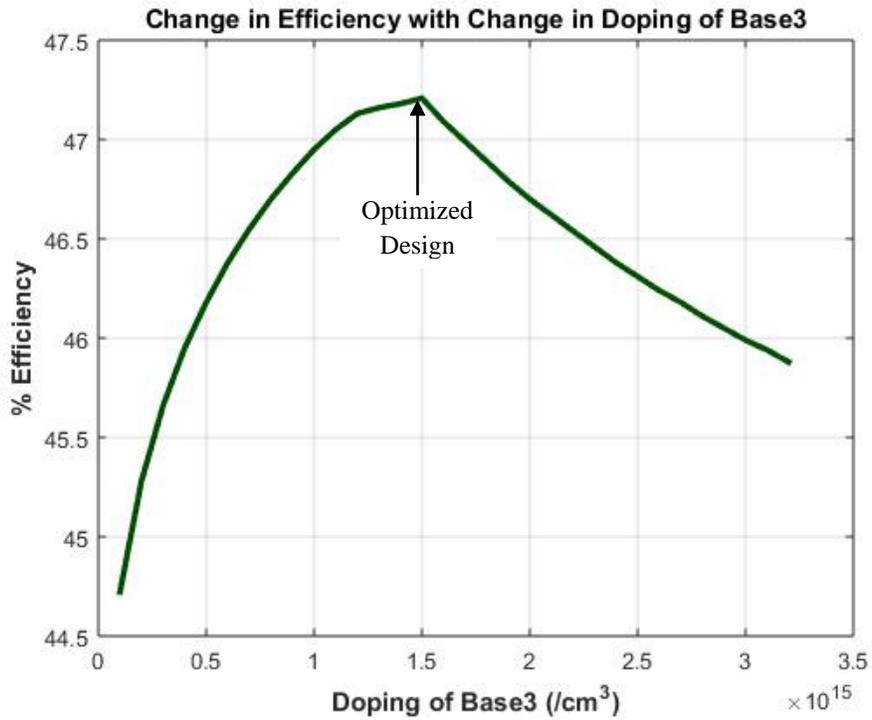

4.4 (f) Efficiency change with change in base3 doping

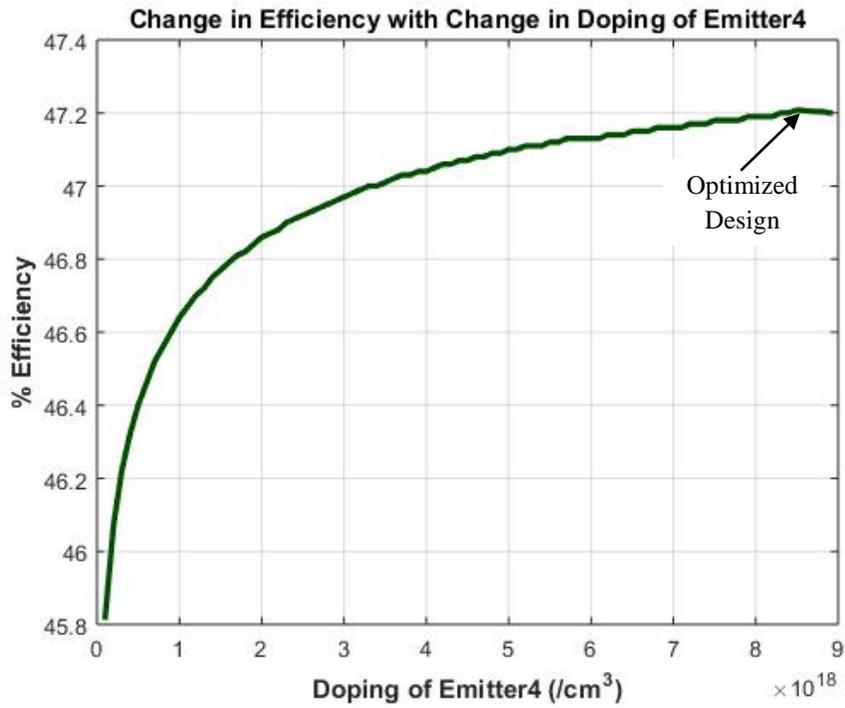

4.4 (g) Efficiency change with change in emitter4 doping

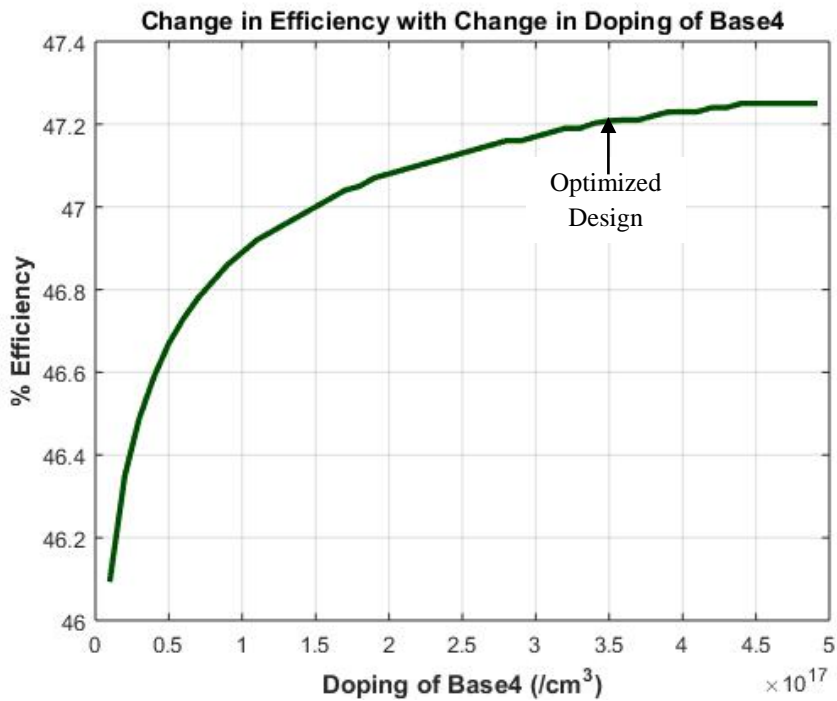

4.4 (h) Efficiency change with change in base4 doping

Fig. 4.4 Change in efficiency with change in doping of a layer, keeping other parameters fixed

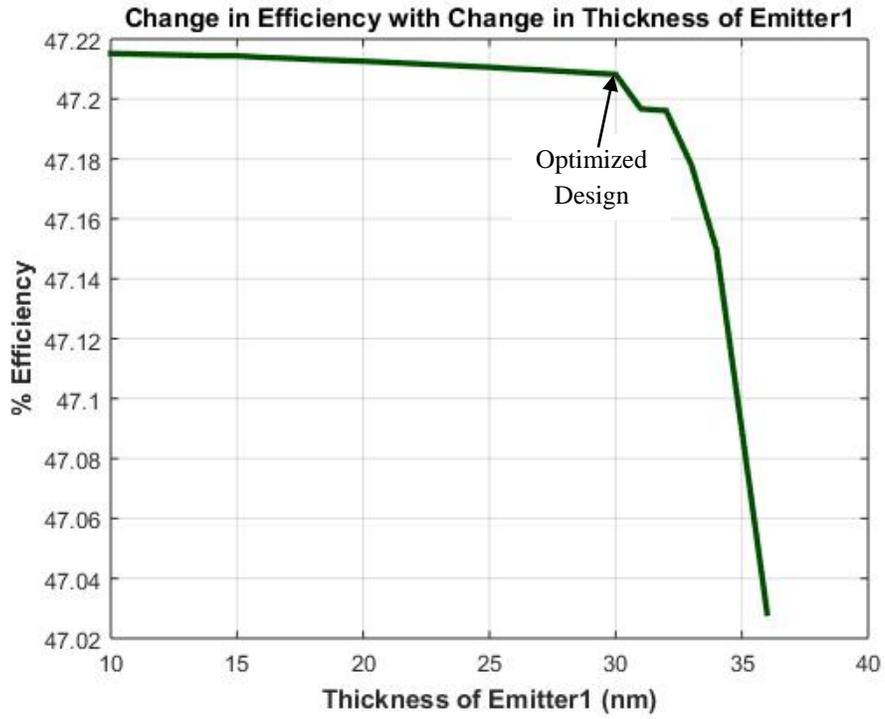

4.5(a) Efficiency change with change in emitter1 thickness

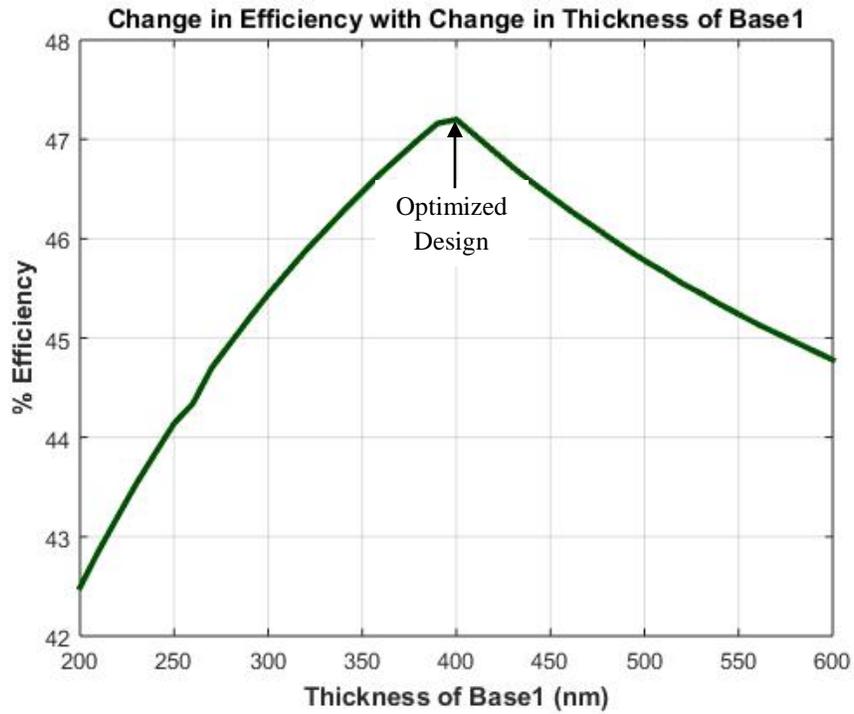

4.5 (b) Efficiency change with change in base1 thickness

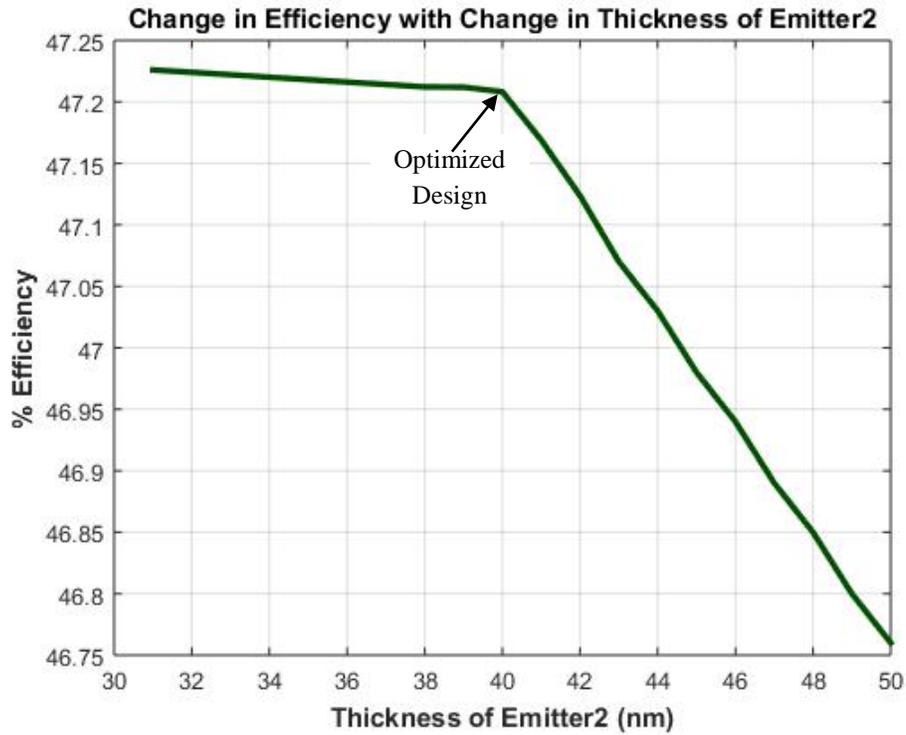

4.5 (c) Efficiency change with change in emitter2 thickness

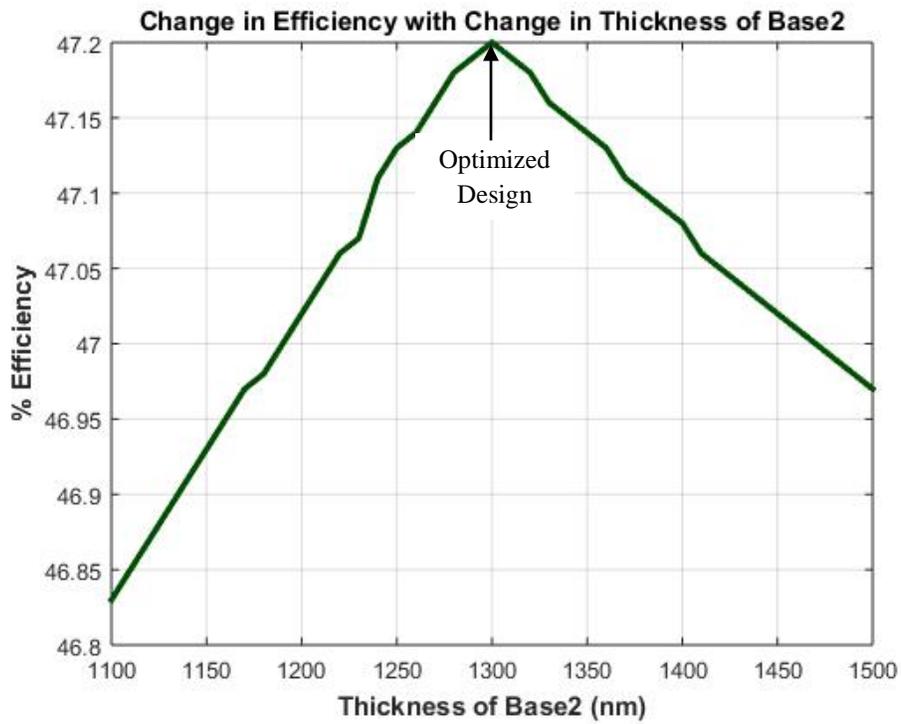

4.5 (d) Efficiency change with change in base2 thickness

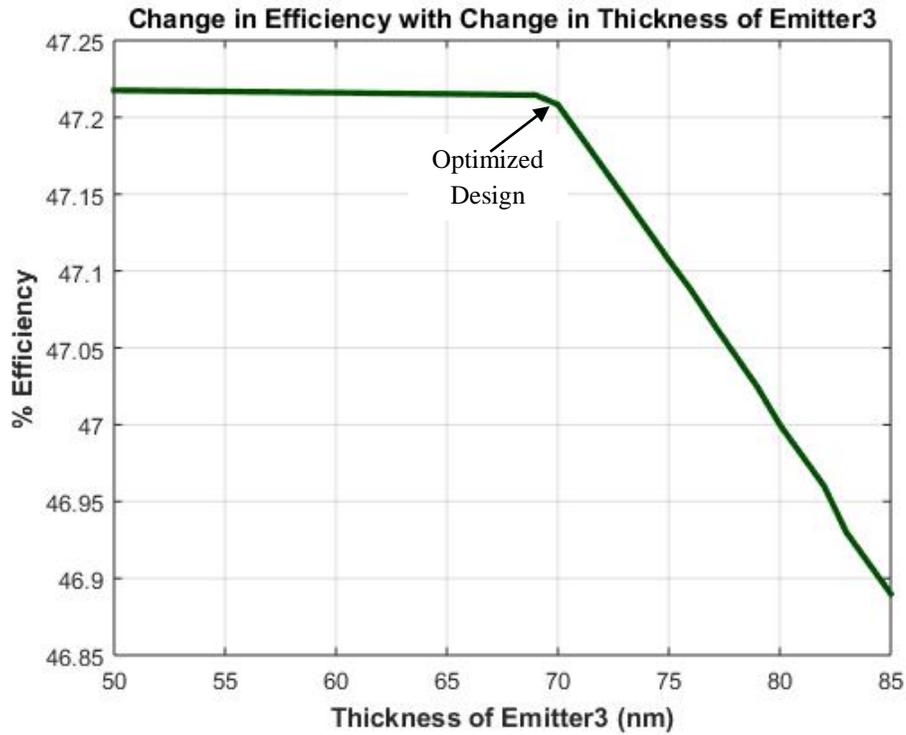

4.5 (e) Efficiency change with change in emitter3 thickness

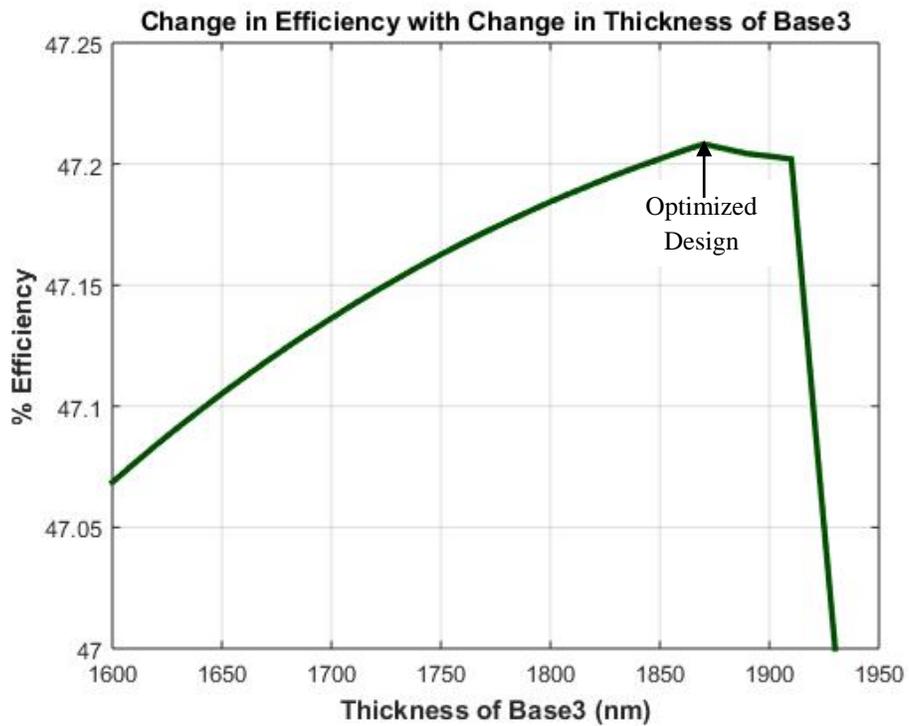

4.5 (f) Efficiency change with change in base3 thickness

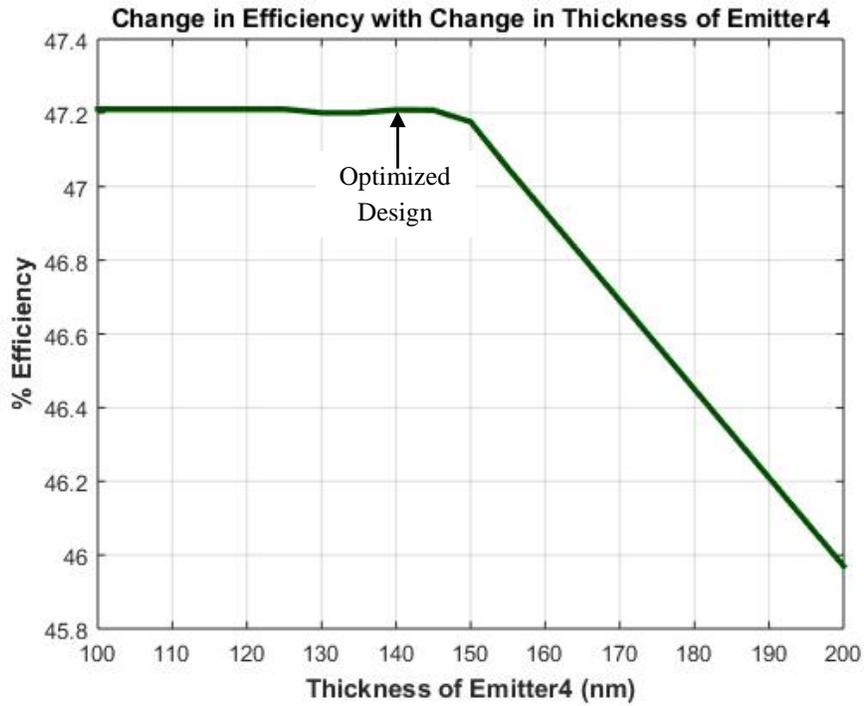

4.5 (g) Efficiency change with change in emitter4 thickness

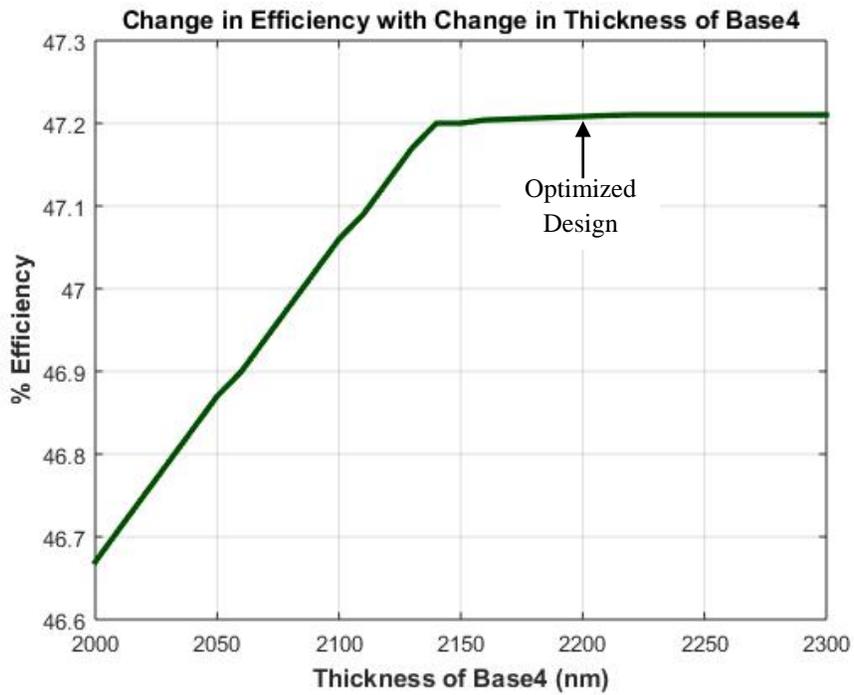

4.5 (h) Efficiency change with change in base4 thickness

Fig. 4.5 Change in efficiency with change in thickness of a layer, keeping other parameters fixed

## 4.6 Analysis of the Optimized Design

### 4.6.1 Quantum Efficiency

The cell was simulated to inspect its quantum efficiency and current density in each of the subcells. We considered global AM 1.5 solar spectrum for the simulation purpose. The internal quantum efficiency (IQE) plot in figure 4 clearly illustrates the absorption properties of the subcells as a function of wavelength. The top subcell, constructed from $In_{.51}Ga_{0.49}P$ showed good exciton (electron-hole pair) generation behavior in the higher frequency visible range. As in figure 4.6, its IQE was more than 90% for green light. GaAs subcell started absorbing when the top subcell was absorbing lesser number of photons. Its IQE was more than 90% in between 500 nm and 828 nm wavelength. It was placed below the top subcell in the stack so that the unabsorbed light can be absorbed by the second subcell. $In_{0.24}Ga_{0.76}As$ showed excellent IQE characteristics in a broad range.

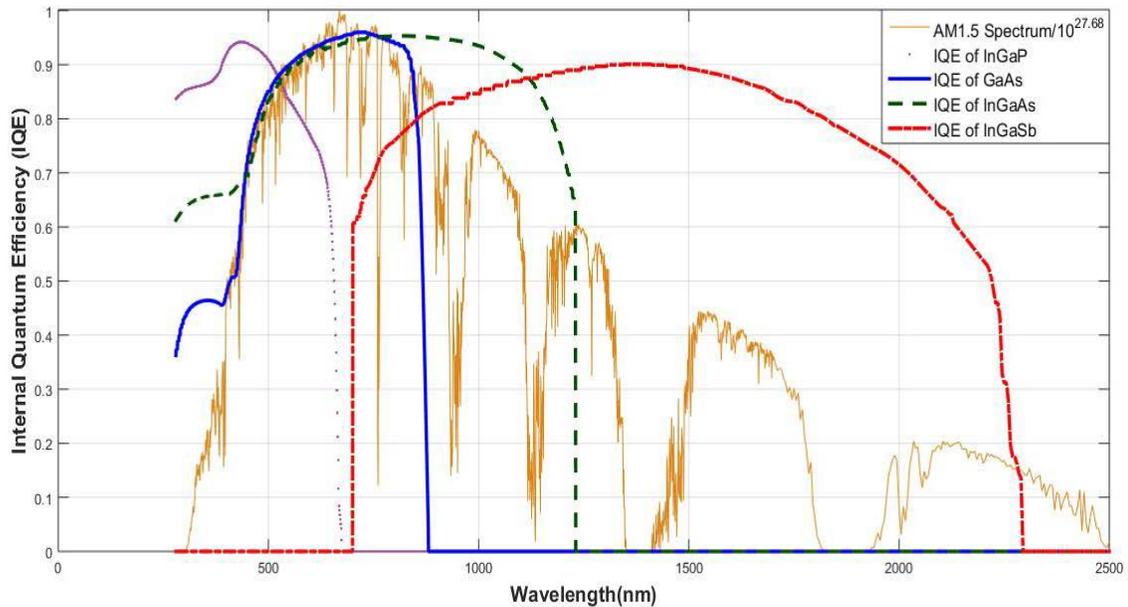

Fig. 4.6 Quantum efficiency plot of the individual junctions in the optimized design

Note that, its IQE value is comparable with GaAs in 500 nm - 828 nm range. If GaAs were not used in the second subcell, the generated current density through $In_{0.24}Ga_{0.76}As$ would be so high that current matching would be very difficult, resulting in lower cell efficiency. The bottom subcell, $In_{0.19}Ga_{0.81}Sb$ absorbed well in the infrared region unlike other subcells. The design ensured the right proportion of light distribution among all the subcells so that generated currents can be easily matched

### 4.6.2  J-V Curve

Figure 4.7 gives an idea about the yield of the corresponding subcells. The top subcell generates the highest voltage 1.4 V with the lowest current of 14.7 mA per 1 $cm^2$ area. The bottom subcell on the contrary gives the lowest voltage (V) of 0.23 V with the highest current density (J) of 50 mA/ $cm^2$. Second and third subcell followed this trend. Thicknesses of the subcells were tuned to attain current matching. In multijunction arrangement, the subcells are connected in series. Therefore, if current is not matched, the excessive current in a subcell, being unable to flow, would be lost as heat and the high temperature could harm the cell further. The J-V curve after current matching is shown in Fig 4.7 (b).

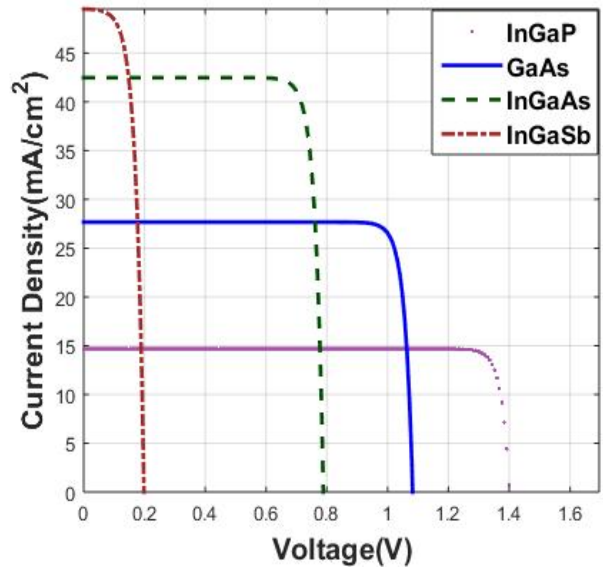

(a)

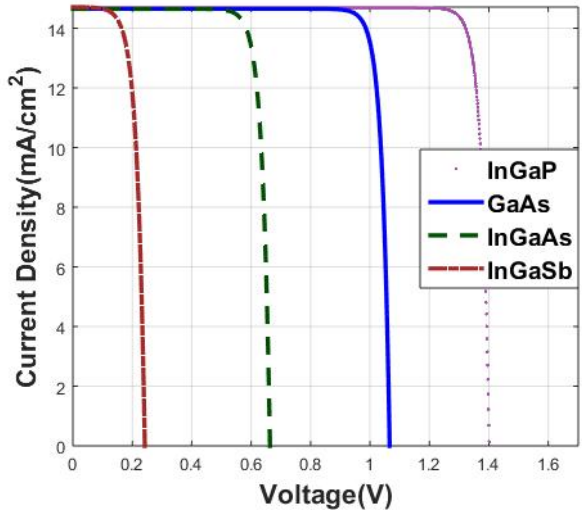

(b)

Fig. 4.7 J-V curve of the cell, (a) before current matching, (b) after current matching

[30] http://www.keysight.com/upload/cmc_upload/All/EE_REF_PROPERTIES_Si_Ga.pdf?&cc=US&lc=eng

[31] http://www.ioffe.ru/SVA/NSM/Semicond/GaAs/electric.html

[32] http://nadirpoint.de/21687_10.pdf

[33] http://www.ioffe.ru/SVA/NSM/Semicond/GaInSb/electric.html#Recombination, accessed on April 30, 2016

[34] http://www.keysight.com/upload/cmc_upload/All/EE_REF_PROPERTIES_Si_Ga.pdf?&cc=US&lc=eng

[35] S. R. Kurtz, P. Faine, and J. Olson, "Modeling of two-junction, series-connected tandem solar cells using top-cell thickness as an adjustable parameter," *Journal of Applied Physics*, vol. 68, no. 4, pp. 1890-1895, 1990.

[36] http://www.cleanroom.byu.edu/OpticalCalc.phtml

[37] http://www.ioffe.ru/SVA/NSM/Semicond/GaInAs/optic.html

[38] Green, Martin A. "Solar cell fill factors: General graph and empirical expressions." *Solid-State Electronics* 24.8 (1981): 788-789.
93

# CHAPTER 5

## ANALYSIS, FUTURE WORK AND CONCLUSION

### 5.1   Realistic Analysis of the Proposed Design

In the last chapter we saw that the proposed design could achieve 47.21% theoretical efficiency, which was calculated using detailed balance method. This method considered ideal conditions i.e. no reflection losses, zero series resistance of subcells and tunnel junctions, 300K temperature and no re-absorption of emitted photons. However practical situations are different from ideal conditions. We will investigate the effects of non-ideal situations in this section.

#### 5.1.1   Non-Ideal Diode

Previously we considered an ideal diode with diode ideality factor of n=1. But in practical case this value is always greater than unity. Change in efficiency of the optimized design was inspected with variation in diode ideality factor value. As illustrated in Fig. 5.1, both the fill factor and efficiency decreases linearly with increase in ideality factor, n. In case of diode ideality factor of 1, efficiency of the proposed optimized design is 47.2082%. For a very bad junction diode with n=2, efficiency drops to 45.5458%. This value is higher than the present record efficiency quadruple junction solar cell in single sun concentration [12]. Most of the practical solar cells have diode ideality factor around 1.2. Considering this value, the efficiency of the proposed cell is 46.6875%.

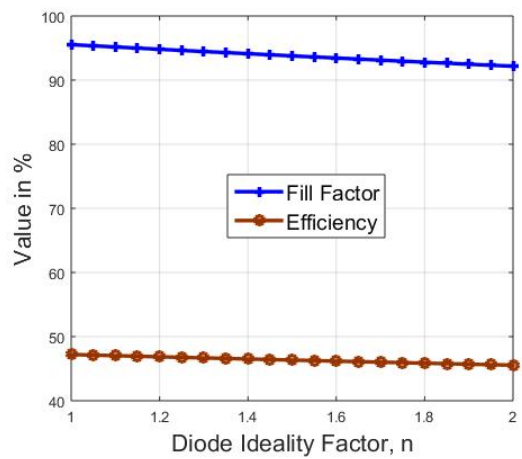

Fig. 5.1 Variation of fill factor and efficiency with diode ideality factor

### 5.1.2 Reflection Loss

Previously we considered no reflection losses and 100% of the incident sunlight were trapped into the solar cell. However, small amount of reflection occurs even after using antireflection coating. The reflectance of present day antireflective coating ranges from 2% to 3% [1, 2]. Double layer $TiO_2$+ $MgF_2$ based double layer antireflection coating is proposed for the proposed quadruple junction solar cell. If we assume 2% reflectance i.e. 98% transmittance, with diode ideality factor of 1.2, the efficiency of the cell becomes 46.6875%*0.98=45.75%, which is still a high value.

### 5.2 Characteristics in Space

### 5.2.1 Existing Design

Incorporating high efficiency solar cells can reduce the burden on fuel consumption of a spacecraft. Therefore it is important to develop high efficiency solar cells. Multijunction solar cells were initially used for space applications only. In space the solar spectrum is of AM0 standard having the flux density of 1353 W/m$^2$ i.e. higher than the AM 1.5 Global spectrums. As depicted in figure 5.2, quantum efficiency values are same as before, because it does not change with illumination.

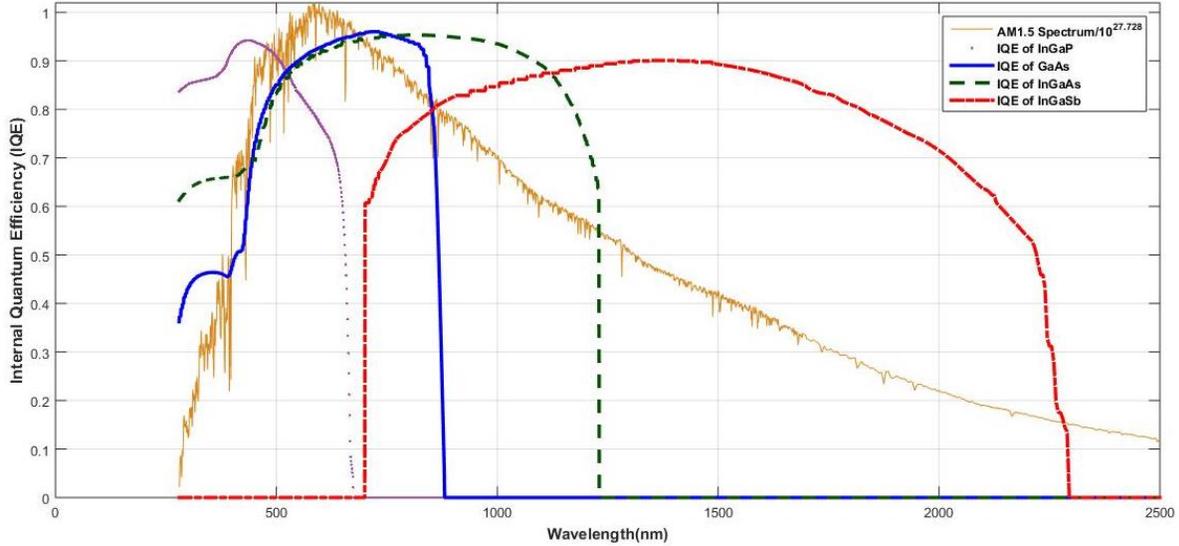

Fig. 5.2 Quantum efficiency of the cell with AM0 spectrum

Table 5.1: Comparison in Performance of the solar cell for AM15G and AM0 Spectrums

| Entity | | AM1.5G Performance | AM0 Performance |
|---|---|---|---|
| Voltage | Subcell 1 | 1.4012 V | 1.4084 |
| | Subcell 2 | 1.0663 V | 1.0691 |
| | Subcell 3 | 0.6635 V | 0.6697 |
| | Subcell 4 | 0.2422 V | 0.2451 |
| Current Density | Subcell 1 | 14.7 mA/cm$^2$ | 19.3 |
| | Subcell 2 | 14.7 mA/cm$^2$ | 16.4 |
| | Subcell 3 | 14.7 mA/cm$^2$ | 18.7 |
| | Subcell 4 | 14.7 mA/cm$^2$ | 24.3 |
| Open Circuit Voltage, $V_{oc}$ | | 3.3731 V | 3.3923 V |
| Short Circuit Current Density, $J_{sc}$ | | 14.7 mA/cm$^2$ | 16.4 mA/cm$^2$ |
| Fill Factor | | 0.9553 | 0.9556 |
| Conversion Efficiency | | 47.2082% | 39.3072% |

Table 5.1 compares the performance of the solar cell for AM1.5 and AM0 spectrums respectively. As expected, photogenerated voltage and current values are higher for AM0 than their AM1.5 counterparts because of higher level of illumination. However, since the input illumination is higher, the efficiency is lower for AM0.

Figure 5.2 depicts the new J-V curve of the solar cell under AM0 illumination which was at first optimized for AM1.5 spectra. The figure suggests that the existing design is inappropriate to use for space applications. As the subcells are in series, the lowest current will flow through the cell. Thermalization loss will occur due to the excess current in other subcells, which will degrade the cell performance further. Therefore, we have to match the current in all the subcells. The design has to be changed for the desired current matching.

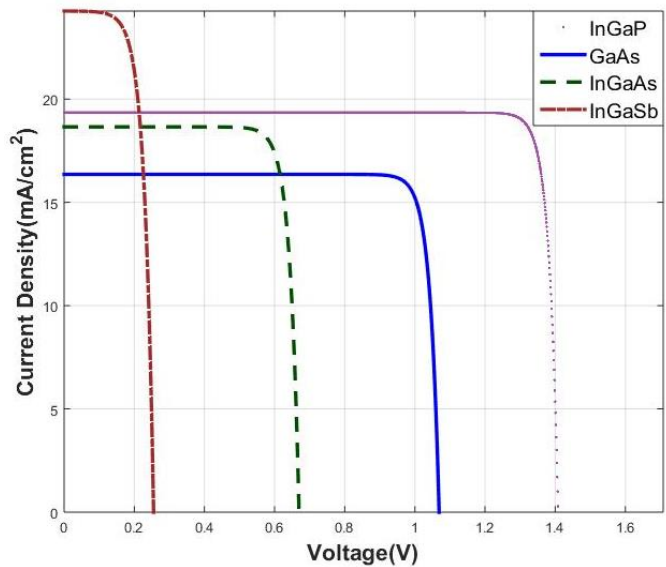

Fig. 5.3 J-V curve in AM0 illumination

### 5.2.2 Modified Design

Thicknesses are changed while keeping the doping levels same; the idea was to use the same material for the new design also. Figure 5.4 illustrates that current density in all the subcells in the new design are matched at 18.5 mA/cm$^2$. The voltage values increased a little bit. The change in the architecture is reflected in table 5.2, which compares the newly optimized design for AM0 with the previous design optimized for AM1.5G. In the modified design thickness values changed for all the layers, except the base of the second subcell. Thus overall thickness is reduced in the new design. In the unmatched case the minimum of the four current densities was $J_{sc.}$ After matching it has increased to 18.5 mA/cm$^2$. Little increase in open circuit voltage and the fill factor is also noticeable. Finally the conversion efficiency has reached to 44.5473%.

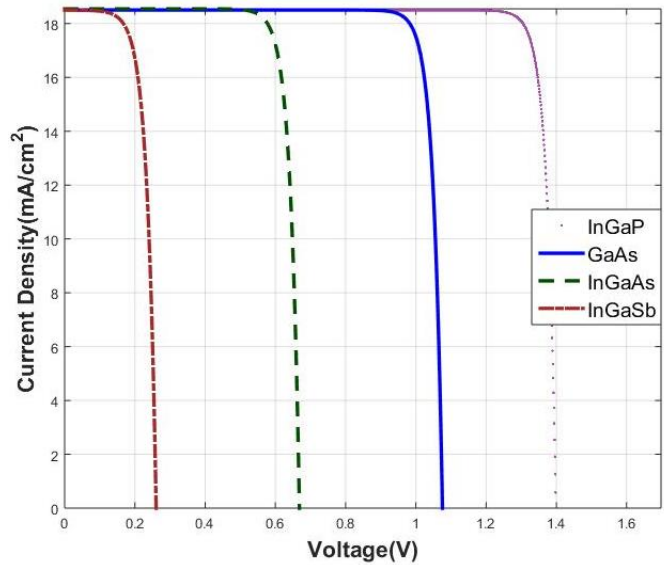

Fig. 5.4 J-V curve of the modified design in AM0 illumination

Table 5.1: Comparison between Previous and Modified Design

| | Parameters | Optimized Design for AM1.5G | Optimized Design for AM0 |
|---|---|---|---|
| Doping Density (/cm$^3$) | Emitter 1 | $8.5 \times 10^{18}$ | $8.5 \times 10^{18}$ |
| | Base 1 | $3.5 \times 10^{17}$ | $3.5 \times 10^{17}$ |
| | Emitter 2 | $3.5 \times 10^{18}$ | $3.5 \times 10^{18}$ |
| | Base 2 | $1.1 \times 10^{15}$ | $1.1 \times 10^{15}$ |
| | Emitter 3 | $8.5 \times 10^{18}$ | $8.5 \times 10^{18}$ |
| | Base 3 | $1.5 \times 10^{16}$ | $1.5 \times 10^{16}$ |
| | Emitter 4 | $8.5 \times 10^{18}$ | $8.5 \times 10^{18}$ |
| | Base 4 | $3.5 \times 10^{17}$ | $3.5 \times 10^{17}$ |
| Thickness (nm) | Emitter 1 | 30 | 10 |
| | Base 1 | 400 | 245 |
| | Emitter 2 | 40 | 20 |
| | Base 2 | 1310 | 1500 |
| | Emitter 3 | 70 | 70 |
| | Base 3 | 1870 | 1700 |
| | Emitter 4 | 140 | 80 |
| | Base 4 | 2200 | 1800 |
| Voltage (V) | Subcell 1 | 1.4084 | 1.3988 |
| | Subcell 2 | 1.0691 | 1.0757 |
| | Subcell 3 | 0.6697 | 0.6685 |
| | Subcell 4 | 0.2451 | 0.2674 |
| $J_{sc}$ | | 16.4 mA/cm$^2$ | 18.5 mA/cm$^2$ |
| Open Circuit Voltage, $V_{oc}$ | | 3.3923 V | 3.4104 |
| Fill Factor (FF) | | 0.9556 | 0.9557 |
| Efficiency | | 39.3072% | 44.5473% |

## 5.3    Future Work

In this research I have designed the quadruple junction solar cell for terrestrial and space power generation (AM 1.5G and AM0 solar spectrums). Future work may include the actual fabrication of the quadruple junction design and finding series resistance of individual subcell layers.

### 5.3.1 Investigating Semiconductor Parameters for Space Atmosphere

Environmental conditions (temperature and pressure) are different in space than on earth. Some extraterrestrial events like solar flare, coronas etc. also occur in space. This surely changes the electrical and optical properties of a material. Thus, while investigating the performance of solar cell in space, the changes in lattice constant, electronic bandgap, surface recombination velocities etc should be considered. I intend to perform that research in future.

### 5.3.1 Fabrication

Since I have achieved noticeably high theoretical light to electrical conversion efficiency in this novel quadruple junction cell structure, I am interested in fabricating the cell and see how much experimental efficiency can be achieved. Inverted metamorphic design will be implemented for MOCVD fabrication.

## 5.4 Conclusion

A quadruple junction solar cell comprising $In_{0.51}Ga_{0.49}P$, GaAs, $In_{0.24}Ga_{0.76}As$ and $In_{0.19}Ga_{0.81}Sb$ subcell layers is proposed in this thesis. This novel III-V combination gives high power conversion efficiency of 47.21% for AM 1.5 global solar spectrum under 1 sun concentration. After careful consideration of important semiconductor parameters such as thicknesses of emitter and base layers, doping concentrations, minority carrier lifetimes and surface recombination velocities, an optimized quadruple junction design has been suggested. Current matching of the subcell layers was ensured to obtain maximum efficiency from the proposed design. Quantum efficiencies were subsequently determined for the matched current density of 14.7 mA/cm$^2$. The proposed quadruple junction solar cell is capable of absorbing and efficiently converting photons from ultraviolet to deep infrared region of the solar radiation spectrum. A modified design has also been proposed for space applications. With a short circuit current density of 18.5 mA/cm$^2$, open circuit voltage of 3.4104 and the fill factor of 0.9557, the power conversion efficiency of the modified quadruple junction design is 44.5473% in space.

# VITA

Mohammad Jobayer Hossain is a former Master's student in the Electrical and Computer Engineering (ECE) Department at Tennessee Technological University (TTU). He completed his B.Sc. in Electronics and Communication Engineering in July 2012 from Khulna University of Engineering and Technology (KUET), Bangladesh. He worked as an ASIC design engineer in a VLSI design farm in Bangladesh for two years after the completion of his bachelor's degree. Since his joining the SOLBAT-TTU Energy Research Laboratory in March 2015, Jobayer has done research in hot carrier solar cells, multijunction solar cell and maximum power point tracking algorithms. He also worked as a Graduate Teaching Assistant for courses such as: capstone design-RADAR project (spring 2015), control system lab (summer 2015) and circuit design lab (fall 2015 & spring 2016). He authored and co-authored several publications in international journals and conference proceedings. His current research interests include optoelectronics, bio-optics and semiconductor materials. Jobayer is a member of SPIE and OSA, and a student member of IEEE.